\documentclass[twocol]{ametsocV6.1}

\usepackage{amsmath,physics}
\usepackage{nicematrix,hhline,diagbox,array}
\usepackage{placeins}
\usepackage{lipsum}
\usepackage{hyperref}
\usepackage{nicefrac}
\usepackage{comment}
\usepackage{soul}

\labelformat{subsection}{\arabic{section}.\alph{subsection}}

\definecolor{Blue}{HTML}{4779c4}
\hypersetup{colorlinks={true},citecolor={Blue},linkcolor={Blue},urlcolor={Blue}}

\newcommand{\fref}[1]{Fig.~\ref{#1}}

\newcommand{\AR}[0]{a} % There is a conflict between the packages ar and nicefrac
\newcommand{\rs}[0]{R^*}
\newcommand{\zs}[0]{{z^*}}
\newcommand{\uc}[0]{U_c}
\newcommand{\ut}[0]{U_t}
\newcommand{\ui}[0]{U_\infty}
\newcommand{\um}[0]{U_M}
\newcommand{\ek}[0]{\epsilon_K}
\newcommand{\ep}[0]{\epsilon_P}
\newcommand{\Ek}[0]{\mathcal{E}_K}
\newcommand{\Ep}[0]{\mathcal{E}_P}
\newcommand{\F}[0]{\mathcal{F}}
\newcommand{\n}[0]{\eta}
\newcommand{\G}[0]{\Gamma}
\newcommand{\oek}[0]{\overline{\epsilon}_K}
\newcommand{\w}[0]{W~kg$^{-1}$}
\newcommand{\esim}[1]{${\ek\sim10^{#1}}$~\w}
\newcommand{\oesim}[1]{${\oek\sim10^{#1}}$~\w}

\newcommand{\cav}[1]{\left\langle#1\right\rangle}

\newcommand{\R}[1]{\textcolor{black}{#1}}

\title{Tidally dominated flows past a three-dimensional topography:\\ Wake vortices, turbulence, and mixing}

\authors{
H. M. Aravind,\aff{a,b} \correspondingauthor{H. M. Aravind, aravind@ucsd.edu}
Pranav Puthan,\aff{a}
Sutanu Sarkar,\aff{a,b}
Sophia Merrifield,\aff{c}
and Eric Terrill\aff{c}
}

\affiliation{
\aff{a}{Mechanical and Aerospace Engineering, University of California San Diego, La Jolla, CA},
\aff{b}{Scripps Institution of Oceanography, La Jolla, CA},
\aff{c}{Coastal Observing R\&D Center, Marine Physical Laboratory, Scripps Institution of Oceanography, La Jolla, CA}
}

\abstract{ %
Oceanic turbulence influences the transport and mixing of freshwater, heat, nutrients, and other biogeochemical tracers.
It also has broader implications for oceanic and atmospheric circulations.
Tides contribute substantially to the mechanically driven turbulent ocean mixing through the internal waves resulting from tide-topography interactions.
Tidal currents also drive turbulent wakes and shear layers when the topography is 3D.
The hypothesis that seamounts are the ``stirring rods'' of the ocean has motivated considerable recent interest in turbulent flow features near 3D topography.
It also motivates the present LES of tidally dominated flows (tidal oscillations superposed on a weaker mean) past an idealized steep seamount.
Complex interactions occur between the topography, the near wake, and previously shed vortices, especially during the tidal phases when the flow direction is reversed.
The topographic wake is shown to be a hotspot for mixing, featuring large dissipation rates in the attached shear layers, hydraulic jet, recirculation region in the near wake, and peripheries of shed vortices.
The majority of the observed dissipation is due to the vertical shear.
Over a tidal cycle, the volume-integrated local dissipation within the wake is at least four times greater than the internal wave flux that may be dissipated elsewhere.
Furthermore, normalized dissipation rates are maximized for the purely tidal setting.
Within the tidal cycle, bulk mixing efficiency ($\eta$) varies substantially and is maximized at $\eta \approx 0.25$ around flow reversals.}

\begin{document}

\maketitle

%%%%%%%%%%%%%%%%%%%%%%%%%%%%%%%%%%%%%%%%%%%%%%%%%%%%%%%%%%%%%%%%%%%%%
% SIGNIFICANCE STATEMENT/CAPSULE SUMMARY
%%%%%%%%%%%%%%%%%%%%%%%%%%%%%%%%%%%%%%%%%%%%%%%%%%%%%%%%%%%%%%%%%%%%%
\statement
Tidal forcing, primarily due to the gravitational attraction from the moon and the sun, constitutes a major source of mechanical energy input into the world ocean. Delineating the energy pathways that ultimately lead to dissipation of the input is crucial to understanding the ocean circulation and mixing, which in turn have broader implications. A significant portion of energy dissipation is attributed to breaking of internal waves generated due to flow-topography interactions. We show that steep seamounts, hypothesized to be stirring rods of the ocean, not only generate internal waves, but also act as local hotspots of energy dissipation. The local dissipation in tidally dominated flows is at least four times greater than the energy transferred to the internal waves, to potentially be dissipated elsewhere.
%TC:endignore

%%%%%%%%%%%%%%%%%%%%%%%%%%%%%%%%%%%%%%%%%%%%%%%%%%%%%%%%%%%%%%%%%%%%%
% MAIN BODY OF PAPER
%%%%%%%%%%%%%%%%%%%%%%%%%%%%%%%%%%%%%%%%%%%%%%%%%%%%%%%%%%%%%%%%%%%%%
%
\section{Introduction}
Turbulence plays a key role in the transport of freshwater, pollutants, nutrients, and other biogeochemical tracers in the ocean~\citep{thorpe2004recent}.
Strong velocity gradients in turbulent motions result in enhanced dissipation rates, which is crucial to the ocean's energy budget.
Specifically, the turbulent dissipation of available potential energy results in irreversible mixing, which plays a central role in maintaining the ocean's stratification~\citep{winters1995available,wunsch2004vertical}.
\citet{munk1998abyssal} estimated that a total of 2.1~TW of dissipation is required to maintain the global abyssal stratification.
While this corresponds to a global average dissipation rate of \esim{-9}, observations reveal that strong turbulent mixing is concentrated in localized hotspots~\citep{waterhouse2014global}, with typical background values in the deep ocean closer to \esim{-10}~\citep{mackinnon2013diapycnal}.

Turbulent dissipation in the ocean interior is thought to be largely driven by the breaking of internal waves, a portion of which originates from the interaction of tidal and subinertial flows with bottom topography~\citep[e.g.,][]{munk1966abyssal,munk1998abyssal,kunze2017internal}.
Internal-wave-related dissipation rates of up to \esim{-7} have been observed across the global ocean \citep[see][for a compilation of observations from moorings, floats, and ADCPS]{waterhouse2014global}.
The mechanisms for the cascade of wave energy leading up to turbulence have been extensively investigated~\citep{mackinnon2013diapycnal,sarkar2017topographic} and parameterized~\citep{mackinnon2017climate}.

%In addition to being a generation mechanism for internal waves,
In addition to wave generation, flow past isolated topographic features can result in localized dissipation near the topography.
For example, flow separation at three-dimensional topography can generate attached shear layers and hydraulic jets, which are sites of enhanced turbulence~\citep{puthan2022wake}.
The laboratory experiments of \citet{hunt1980experiments} and \citet{castro1983stratified} revealed that wake eddies are shed in the lee of dynamically tall obstacles when the steady inflow does not have sufficient kinetic energy to overcome the potential energy barrier imposed by the stratification.
Vertically \R{coupled} wake vortices in the lee of three-dimensional topography exposed to a steady current were extensively investigated through numerical simulations, using hydrostatic models~\citep{dong2007island,perfect2018vortex,perfect2020energetics1,srinivasan2021high,jagannathan2021boundary} and non-hydrostatic large eddy simulations~\citep{puthan2020wake,liu2024effect}, and were shown to be regions of enhanced dissipation. %~\citep{,}.

Wake eddies in the lee of islands and submarine ridges were revealed by a number of recent observational studies, and dissipation rates as high as \esim{-5} were observed in the wake~\citep{chang2013kuroshio,chang2019observations,zeiden2019glider,mackinnon2019eddy,johnston2019energy,st2019turbulence,voet2020topographic,wynne2022measurements}.
The wake dynamics in these oceanographic settings depend strongly on the flow impinging on the topography.
Specifically, tidal oscillations can result in significant changes to the wake behavior, compared to a steady inflow.
For example, the large eddy simulations (LESs) of \citet{puthan2021tidal} revealed tidal synchronization of the wake eddies.
\citet{puthan2022wake} observed dissipation rates of up to \esim{-5} in the turbulent near wake, and noted that synchronization in tidally modulated flows occurred only if the amplitude of the tidal current is at least 20\% of the mean flow.

The tidal component of the flow interacting with topography can be comparable to or even larger in velocity than the mean current.
Such cases of flow-topography interaction occur in the deep ocean and have been observationally studied \citep{dale2015tidal,voet2020topographic} from the perspective of nonlinear internal waves and turbulence on the slopes.
The maximum flow velocity was not large (0.7~ms$^{-1}$ in \citealp{dale2015tidal} and 0.2~ms$^{-1}$ in \citealp{voet2020topographic}), but the topographies had steep slopes and modest widths $\sim1$~km, so that the tidal excursion number was $\sim1$. In contrast, coastal regions subject to strong barotropic tides can exhibit peak flows exceeding 1~ms$^{-1}$ \citep{lu2000turbulence}. Motivated by the flow regime of these examples, we perform an LES study of {\em tidally dominated} flows where the relative magnitude of the mean is systematically varied from 0 (pure tide) to 1 and the tidal excursion number exceeds unity.  We will show that the turbulent wake dynamics in this regime of tidal domination are substantially different and constitute a different regime from previous LES work with tides.  Although the focus is on wake eddies and turbulence, the internal wave flux (a surrogate for remote dissipation) will be compared to the turbulent dissipation in the wake.

The computational model used for the LES is discussed in \S~\ref{ssec:model}.
The key non-dimensional parameters and other quantities of interest are discussed in \S~\ref{ssec:defns}.
The values of the physical parameters, and corresponding non-dimensional groups are presented in \S~\ref{ssec:params}.
% The vortex dynamics in the wake within a single tidal cycle is discussed using three representative cases of tidally dominated flows in \S~\ref{ssec:dynamics}, and \S~\ref{ssec:instantaneous} presents the dissipation rates observed in the simulations.
% Phase averaging of data over ten tidal cycles, after the system has achieved statistical stationarity, allows an investigation of dissipation and mixing statistics, presented in \S~\ref{ssec:phase-averaged}.
\R{The vortex dynamics in the wake within a single tidal cycle is discussed using three representative cases of tidally dominated flows in \S~\ref{ssec:dynamics}, wherein the salient features in the streamwise velocity and the vertical vorticity are highlighted.
The link between observed flow features and the dissipation rate of kinetic energy is then discussed in \S~\ref{ssec:instantaneous}, and the role of vertical shear in the system is delineated.
Tidal-phase-averaged dissipation rate is then investigated in \S~\ref{ssec:phase-averaged}, to obtain statistics of the volume-integrated dissipation rate in the wake (``wake dissipation''), which is compared against the area-integrated internal wave flux (``wave dissipation'').
The dissipation of available potential energy is also investigated, and the mixing efficiency within the wake is quantified as a function of the tidal phase.}
Finally, \S~\ref{sec:disc} presents a discussion of the key findings and their implications.

\section{Theory and methods}
\label{sec:theory}
\subsection{Model Setup}
\label{ssec:model}

%TC:ignore
\begin{figure*}[!t]
    \centering
    \includegraphics[scale=0.4]{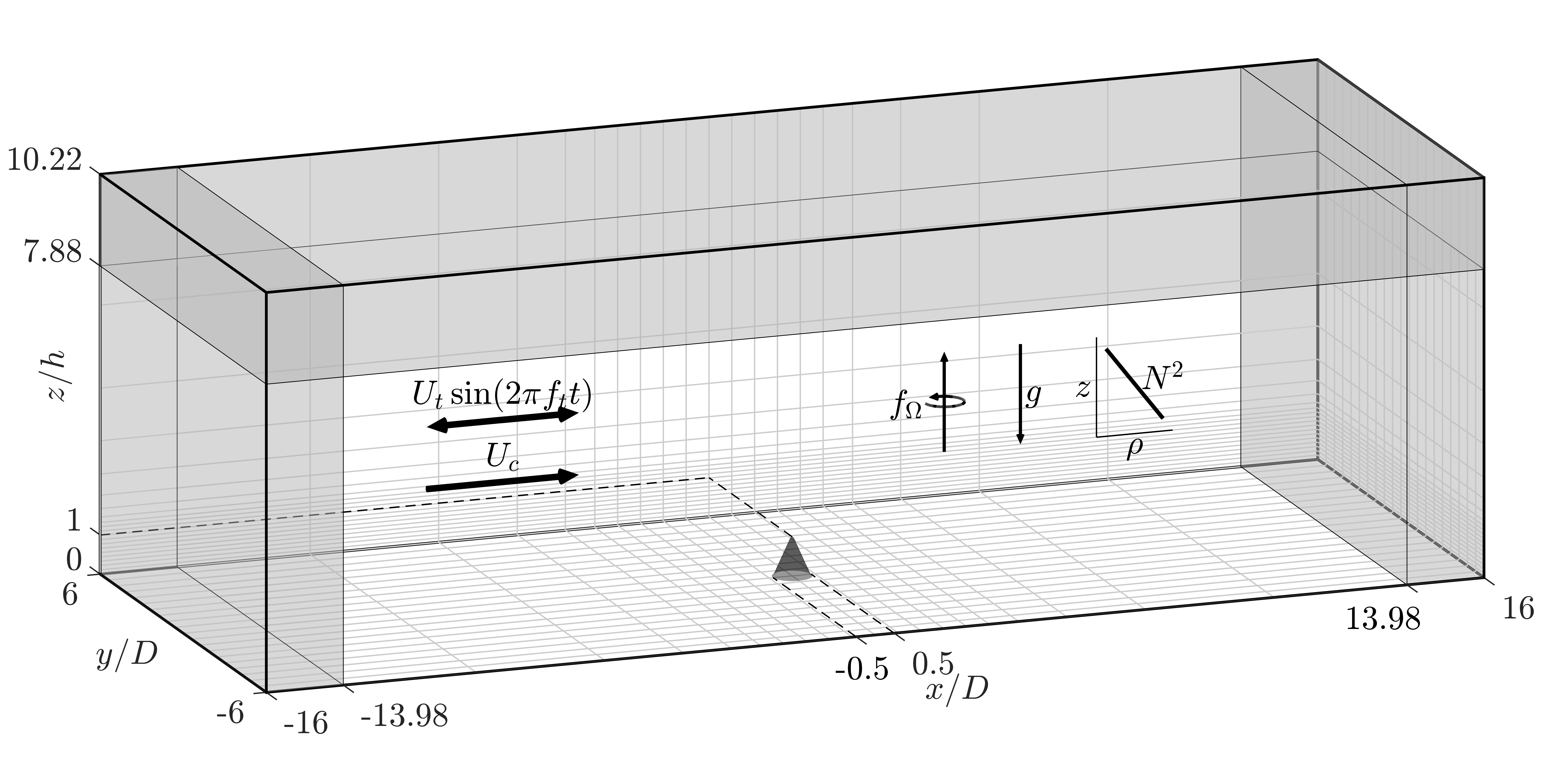}
    \caption{A schematic of the problem setup.
    A description of the various symbols is in \S~\ref{sec:theory}.
    The gray-shaded regions denote the three sponge layers near the western ($x/D\approx-16$), eastern ($x/D\approx16$), and top ($z/h\approx10$) boundaries.
    The light gray lines correspond to cell edges in the computational grid --- a line is plotted every 128, 128, and 20 edges in the $x$-, $y$-, and $z$-directions, respectively.}
    \label{fig:schematic}
\end{figure*}
%TC:endignore

The computational model (see \fref{fig:schematic} for a schematic) simulates the interaction of an unsteady barotropic current with an idealized three-dimensional (3D) topography, in a linearly stratified setting with background rotation.
The unsteady free-stream $\ui(t)=\uc + \ut \sin(2\pi f_t t)$ is modeled as a superposition of a mean current $\uc$ and an oscillating tidal component $\ut \sin(2\pi f_t t)$, and the conical seamount has height $h$ and base diameter $D$.
The buoyancy frequency $N$ is a constant, and the Coriolis parameter is $2\pi f_\Omega$.

The numerical model solves the three-dimensional Navier-Stokes equations filtered for large eddy simulations, under the Boussinesq approximation in a rotating framework, expressed using the index notation as:
\begin{equation}
    \pdv{u_i}{x_i} = 0,
\end{equation}
\begin{equation}
\begin{aligned}
    \pdv{u_i}{t} &+ \pdv{(u_i u_j)}{x_j} - \varepsilon_{ij3} (2\pi f_\Omega)(u_j - \ui\delta_{j1}) \\&= -\frac{1}{\rho_0} \pdv{p^*}{x_i} - \frac{g\rho^*}{\rho_0} \delta_{i3} + F_\infty(t) \delta_{i1} + \pdv{\tau_{ij}}{x_j},
\end{aligned}
\end{equation}
\begin{equation}
    \pdv{\rho}{t} + \pdv{(u_j \rho)}{x_j} = \pdv{\Lambda_j}{x_j}.
\end{equation}

Here, $u_i$ represents the filtered velocity component in the $i^\mathrm{th}$ direction, with $i=1,2,3$ corresponding to the streamwise/zonal ($x$), spanwise/meridional ($y$), and vertical ($z$) directions, respectively.
The forcing term
\begin{equation}
    F_\infty(t)=\frac{\mathrm{d}U_\infty}{\mathrm{d}t} = 2\pi f_t \ut \cos(2\pi f_t t)
\end{equation}
sustains the barotropic tidal oscillations in the system.
The total density $\rho$ is modeled as a sum of a reference density $\rho_0$, a vertically varying $\rho_r$, and a deviation $\rho^*$:
\begin{equation}
    \rho = \rho_0 + \rho_r(z) + \rho^*(x,y,z,t) = \rho_b(z) + \rho^*(x,y,z,t).
\end{equation}
The total pressure $p$ is a sum of the ambient pressure $p_a$, a hydrostatic component $p_b$ that balances the contribution from $\rho_b$, a meridionally varying $p_g$ that is in geostrophic balance with the barotropic $U_\infty(t)$, and a deviation $p^*$:
\begin{equation}
    p = p_a + p_b(z) + p_g(y,t) + p^*(x,y,z,t).
\end{equation}
The stress tensor \R{$\tau_{ij} = (\nu + \nu_{SGS})\left({\partial u_i}\big/{\partial x_j}\right)$} incorporates both the molecular viscous stress and the subgrid-scale (SGS) stress, calculated using the \R{Wall-Adapting Local Eddy-viscosity (WALE)} model \citep{nicoud1999subgrid}:\R{
\begin{gather}
\nu_{SGS} = (C_w\Delta)^2 \frac{\left(\mathscr{S}^d_{ij}\mathscr{S}^d_{ij}\right)^{3/2}}{\left(S_{ij}S_{ij}\right)^{5/2} + \left(\mathscr{S}^d_{ij}\mathscr{S}^d_{ij}\right)^{5/4}}, \mathrm{~where~}
\\
\mathscr{S}^d_{ij} = S_{ik}S_{kj} + \Omega_{ik}\Omega_{kj} - \frac{1}{3}\delta_{ij}\left( S_{mn}S_{mn} - \Omega_{mn}\Omega_{mn} \right),\\
S_{ij} = \frac{1}{2}\left(\frac{\partial u_i}{\partial x_j}+\frac{\partial u_j}{\partial x_i}\right),\hspace{1em}
\Omega_{ij} = \frac{1}{2}\left(\frac{\partial u_i}{\partial x_j}-\frac{\partial u_j}{\partial x_i}\right),
\end{gather}
$C_w=0.5$ is a constant, and $\Delta = \left(\Delta x\Delta y\Delta z\right)^{1/3}$, where $\Delta x$, $\Delta y$, $\Delta z$ are local grid sizes in the $x$-, $y$- and $z$- directions, respectively.
The model accounts for the role of small-scale eddies in turbulent dissipation and ensures $\nu_{SGS}$ asymptotes to zero at a wall, by construction.
Similarly, the scalar flux term $\Lambda_j$ is given by
$\Lambda_j = (\kappa + \kappa_{SGS}) \left({\partial\rho}\big/{\partial x_j}\right)$, with $\kappa_{SGS}=\nu_{SGS}$.}

The unsteady, three-dimensional governing equations are solved numerically on a staggered grid that spans $x/D\in[-16,16]$ in the streamwise direction, $y/D\in[-6,6]$ in the spanwise direction, and $z/h\in[0,10.22]$ in the vertical.
The domain is discretized on a $2562\times2562\times402$ grid with a total of 2.7 billion points.
A uniform grid is used near the topography, within $x/D\in[-3,3]$, $y/D\in[-3,3]$ and $z/h\in[0,1.32]$, and a mild grid stretching (by factors of 1.0038 and 1.023, in the $x$- and $z$-directions, respectively) away from the topography.
The topography is modeled using the immersed boundary method of \cite{YangB_JCP:2006}, with no-slip and adiabatic boundary conditions on its surface.
Sponge layers~\citep{taylorS_JFM:2007} that damp the flow to the free-stream conditions are placed at the top and the streamwise boundaries, and the domain is periodic in the spanwise direction.
Spatial derivatives are discretized using a second-order central difference, and the time integration is performed using a semi-implicit, low-storage Runge-Kutta (RK-3)/Crank-Nicholson scheme \citep{Fletcher:91, Bewely2007}.

\subsection{Key parameters and definitions}
\label{ssec:defns}
Important physical parameters that govern the dynamics of the selected problem include topographic geometry,  height $h$ and diameter $D$ of the conical seamount; the current, its mean velocity $\uc$, tidal velocity amplitude $\ut$ and the tidal frequency $f_t$; the background, the Coriolis frequency $f_\Omega$ and the buoyancy frequency $N$.
Associated non-dimensional parameters and their physical significance are outlined below.
The aspect ratio of the conical seamount, $\AR=h/D$, determines the slope angle $\beta = \tan^{-1}\left( {h}/ (D/2) \right)$, which plays a key role in the internal wave dynamics in the system.
For example, the reflection of incident internal waves propagating at an angle $\alpha$ depends on the slope criticality $\gamma=\tan(\beta)/\tan(\alpha)$.
The value of $\gamma$ also determines turbulence during internal tide generation, e.g., resonant boundary layer response at near-critical slope with $\gamma \approx 1$ \citep{gayen2011direct}, breaking lee waves at supercritical ($\gamma > 1$) ridges~\citep{legg2008internal,alford2014breaking}, and wave-beam boundary layers  \citep{gayen2011boundary} at $\gamma \geq 1$ slopes.

The velocity ratio $\rs=\uc/\ut$, a measure of tidal strength relative to the mean, is the key non-dimensional parameter of interest in this study.
To understand tidally dominated flow, we investigate the regime $\rs\le1$.
\cite{puthan2022wake} has already considered the other regime ($R=1/\rs<1$) of tidally modulated flow.

%TC:ignore
\begin{table*}[!t]
\centering
\begin{NiceTabular}{|W{c}{3.0cm}||W{c}{2.0cm}|W{c}{2.0cm}|W{c}{2.0cm}|W{c}{2.0cm}|W{c}{2.0cm}|}%[hvlines]
    \CodeBefore
    % \cellcolor[HTML]{DCDCDC}{1-,-1,-2}
    % \cellcolor[RGB]{220,220,220}{1-,-1}
    % Removed shading according to JPO guidelines; added double lines instead
    \Body
    \Hline
    $\rs=\nicefrac{\uc}{\ut}$ & 1.00 & 0.75 & 0.50 & 0.25 & 0.00 \\ \hline\hline
    $Ro=\nicefrac{\um}{2\pi f_\Omega D}$ & 10.60 & 9.27 & 7.95 & 6.62 & 5.30 \\ \hline
    $Fr_h=\nicefrac{\um}{Nh}$ & 0.30 & 0.26 & 0.23 & 0.19 & 0.15 \\ \hline
    $Ex_c=\nicefrac{\uc}{2f_tD}$ & 4.49 & 3.37 & 2.24 & 1.12 & 0.00 \\ \hline
    $Ex^-=Ex_c-Ex_t$ & 1.63 & 0.51 & -0.61 & -1.74 & -2.86 \\ \hline
    $f^*=f_{vs,c}/f_t$ & 2.35 & 1.75 & 1.16 & 0.56 & 0.00 \\ \hline
     $Re=\nicefrac{\um D}{\nu}$ & 20000 & 17500 & 15000 & 12500 & 10000 \\ \hline
    \Hline
\end{NiceTabular}
\vspace{0.5em}
\caption{Important non-dimensional groups. Note that the aspect ratio of the seamount ${\AR=h/D=0.3}$, tidal excursion number ${Ex_t=\nicefrac{\ut}{\pi f_tD}=2.86}$\R{, and the Burger number $Bu =(\nicefrac{Ro}{Fr_h})^2 = (\nicefrac{Nh}{2\pi f_\Omega D})^2=1253.5$,} for all five cases considered.}
\label{tab:params}
\end{table*}
%TC:endignore

The tidal excursion number $Ex_t=\ut/\pi f_t D$, which is a non-dimensional distance traveled by a fluid particle due to tidal flow over half the tidal cycle (tidal excursion length), is also of interest.
\cite{winters2013response}, for example, discusses how a modified tidal excursion number is a measure of the relative importance of inertial and time-dependent effects near the topography.
An excursion number can also be defined based on the mean velocity of a tidal current, as $Ex_c=\uc/2f_tD$.

A useful parameter for the organization of wake vortices when $\ut\gtrsim\uc$ is the excursion number for the second half of the tidal cycle ($\pi < \phi < 2 \pi$),
\begin{equation} \label{eq:exm}
    Ex^-(\rs)=Ex_c-Ex_t= \frac{\ut}{2f_tD}\left(\rs-\frac{2}{\pi}\right).
\end{equation}
Eq.~\eqref{eq:exm} implies that the net particle advection during the tidal half-cycle when the tidal current opposes the mean is negative ($Ex^-<0$) when $\rs<2 /\pi$.
The magnitude of $Ex^-$ quantifies relative displacements by the mean and tidal component during $\pi < \phi < 2 \pi$, and influences the vortices in the vicinity of the topography, as will be shown later.

Another parameter of interest is the frequency ratio $f^*=f_{vs,c}/f_t$~\citep{puthan2021tidal}, where $f_{vs,c}=St_c\uc/D$ is an estimate of the natural vortex shedding frequency in the absence of tidal oscillations.
The Strouhal number ${St_c = 0.273 - 1.11Re_c^{-0.5} + 0.482Re_c^{-1}}$ is calculated with the $\uc$-based Reynolds number, using the ${St_c-Re_c}$ relationship for a cylinder wake proposed by \cite{williamson1998series}.

In stratified flows with a velocity scale $U$, the fluid particles have sufficient kinetic energy to overcome vertical displacements of up to $U/N$.
The topographic Froude number is a ratio of this vertical length scale to the height of the topography, and indicates the depth below which the upstream flow is blocked~\citep{baines2022topographic,winters2014topographic}.
Here, we define the Froude number as $Fr_h=\um/Nh$, where $\um=\uc+\ut$ is the free-stream velocity at peak tidal flow.
The topography is {\em dynamically tall} if $Fr_h<1$, and nonlinear effects can occur.

The Rossby number $Ro=\um/2\pi f_\Omega D$ quantifies the importance of the background rotation in the system, where $2\pi f_\Omega$ is the Coriolis parameter.
The earth's rotation plays a critical role in the mesoscale regime, when $Ro\lesssim10^0$.
For tidally dominated flows, even if $Ro\gtrsim10^0$, background rotation can play a significant role during parts of the tidal cycle when the tidal flow is opposite to the mean current.

The Reynolds number $Re=\um D/\nu$ is also defined based on the velocity at peak tidal flow.
To understand the dissipation and mixing in the wake of the three-dimensional topography, we investigate $Re\sim10^4$ wherein a turbulent near wake is observed downstream of the seamount~\citep{puthan2022wake}.
% Note, that we use ${\R{\xi}\sim10^{n}}$, where $n$ is an integer \R{and $\R{\xi}$ is a quantity of interest}, iff ${10^{-0.5}\le\R{\xi}\times10^{-n}\le10^{0.5}}$ when investigating the magnitudes of the various quantities presented in this study.
\R{Additionally, the non-dimensional numbers presented in Table~\ref{tab:params} are to be associated with the bulk dynamics over an entire tidal cycle. The instantaneous, local magnitudes of these parameters can span a wide range of values.}

The dissipation rates of kinetic energy per unit mass $\ek$ and available potential energy per unit mass $\ep$ are calculated as
\begin{align}
    \ek&=2(\nu + \nu_{SGS})S_{ij}S_{ij}\mathrm{,~and~}\\
    \ep&=2(\kappa+\kappa_{SGS}) \left(\frac{g^2}{2\rho_0^2N^2}\right) \frac{\partial\rho^*}{\partial x_j}\frac{\partial\rho^*}{\partial x_j},\label{eq:ep}
\end{align}
where $\rho^*$ is the deviation from the linear background density profile~\citep{kang2010calculation}.
\R{Note, that the definition in Eq.~\eqref{eq:ep} is valid for the dissipation rate of available potential energy only for a linear background stratification, which is the case in this study.}
To obtain statistically robust measures of the ``wake dissipation,'' volume integrals of the tidal-phase-averaged dissipation rates are computed.
\R{The `tidal phase average' (or `phase average') $\overline{\xi}(\phi)$ of a quantity $\xi(t)$ at a tidal phase $0\le\phi<2\pi$ is defined as
\begin{equation}
    \overline{\xi}(\phi)=\sum_{n=n_1}^{n_2}\xi\left(t=\frac{(n-1)+(\phi/2\pi)}{f_t}\right)\Bigg/(n_2-n_1+1).
\end{equation}
For our analysis, we allowed the simulations to spin up over nine tidal cycles, and used the flow fields from the next ten cycles for phase averaging, i.e., $n_1=10$ and $n_2=19$.}

The volume integrals used to obtain the wake dissipation rates $\Ek$ and $\Ep$ are computed using a trapezoidal rule on the data from four horizontal transects at $z/h\in\{0.10,0.25,0.5,0.75\}$ and a midpoint rule on the three vertical transects at $y/D\in\{-0.375,0,0.375\}$.
The integration in the vertical direction is performed from $z/h=0^+\equiv0.025$,  thus excluding the thin bottom viscous layer, up to $z/h=1.2$.
%, to account for the vertical growth of the wake.
Horizontally, the volume integration spans the entire spanwise extent of the domain ($12D$) and  $\pm 10D$ in the streamwise direction.
This ``wake region,'' within which the wake dissipation rates are calculated, encloses a volume of approximately 10.6~km$^3$.

The wake dissipation is then compared against the ``wave dissipation,'' i.e., kinetic energy transported away from the seamount by the area-integrated internal wave flux $\F$~\citep{perfect2020energetics2,puthan2022wake}, to potentially be dissipated elsewhere.
The quantity $\F$\R{, defined as
\begin{equation}
    \F=\int_{-6D}^{6D}\int_{-10D}^{10D}\left.\overline{p^*w}\right|_{z=1.2h}~\mathrm{d}x\mathrm{d}y,
\end{equation}}
is computed as an integral over the 60~km$^2$ area of the top of the subdomain within which the wake dissipation is calculated.
Finally, the bulk mixing efficiency $\n=\Ep/(\Ep+\Ek)$~\citep{scotti2014diagnosing,puthan2019energetics} is computed to quantify the extent of mixing occurring in the system, as a function of the tidal phase.
The mixing efficiency is related to the mixing coefficient $\G\simeq\Ep/\Ek=\n/(1-\n)$~\citep{caulfield2021layering}, which is usually used to estimate diffusivities from available dissipation measurements.
Although \citet{osborn1980estimates} postulated $\G\le0.2$ in a statistically steady state in the ocean, the inequality is often neglected in practice.

\subsection{Selection of simulations}
\label{ssec:params}
Table~\ref{tab:params} presents the values of the different non-dimensional groups discussed above, in \S~\ref{ssec:params}.
The Froude number and the aspect ratio of the topography that interacts with the simulated barotropic flow in a linear stratification are representative of abyssal seamounts that act as `stirring rods' in the ocean.
Since our study is motivated by the variability in boundary currents experienced by bottom topography, we investigate the effect of varying the velocity ratio $R^*$ by fixing $\ut$ to a constant value and varying the magnitude of $\uc$.
To investigate tidally dominated flows, we consider five values of the velocity ratio, $\rs\in\{1.00,0.75,0.50,0.25,0.00\}$, listed in decreasing order of the magnitude of $\uc$.

In dimensional terms, five values of $\uc$ are considered between $0.1$ and $0$~m~s$^{-1}$.
The tidal flow has an amplitude $\ut=0.1$~m~s$^{-1}$ and a frequency $f_t=2.23\times10^{-5}$~s$^{-1}$, which corresponds to the $M_2$ tidal component.
The idealized seamount in the simulations has a height of $h=150$~m and a base diameter of $D=500$~m.
The buoyancy frequency of $N=4.44\times10^{-3}$~s$^{-1}$ associated with the linear stratification results in $Fr_h\le0.3$, and $2\pi f_\Omega=3.77\times10^{-5}$~rad~s$^{-1}$ is the Coriolis parameter for a $15^\circ$~N Latitude.
%The kinematic viscosity $\nu=5\times10^{-3}$~m$^2$~s$^{-1}$ corresponds to a high enough
The Reynolds number is sufficiently large to establish a turbulent wake.

% \section{Results}
% We start with a discussion of the tidal phasing of the flow in \S~\ref{ssec:dynamics} where three representative cases ($\rs= \uc/\ut = 1.00, 0.50$ and $0.00$) are contrasted and the salient features in the streamwise velocity and the vertical vorticity ($\omega_z$) are highlighted.
% The link between observed flow features and the dissipation rate ($\ek$)
% %of kinetic energy per unit mass
% is then discussed in \S~\ref{ssec:instantaneous}, and the role of vertical shear in the system is delineated.
% Tidal-phase-averaged dissipation rate ($\bar{\epsilon}_K$) is then investigated in \S~\ref{ssec:phase-averaged}, to obtain statistics of the volume-integrated dissipation rate in the wake (``wake dissipation'' $\Ek$), which is compared against the area-integrated internal wave flux (``wave dissipation'' $\F$).
% The dissipation of available potential energy  ($\Ep$) is also investigated and the mixing efficiency ($\n=\Ep/(\Ep+\Ek)$) within the wake is quantified as a function of the tidal phase.
% Overall, despite some cycle-to-cycle variability in the flow, the phase-averaged values provide a robust description of turbulent dissipation and mixing in the system.

\section{Temporal dynamics: turbulent wake, \R{vertically coupled} vortices, and lee waves}
\label{ssec:dynamics}

%TC:ignore
\begin{figure*}[!t]
    \centering
    \includegraphics[scale=0.714]{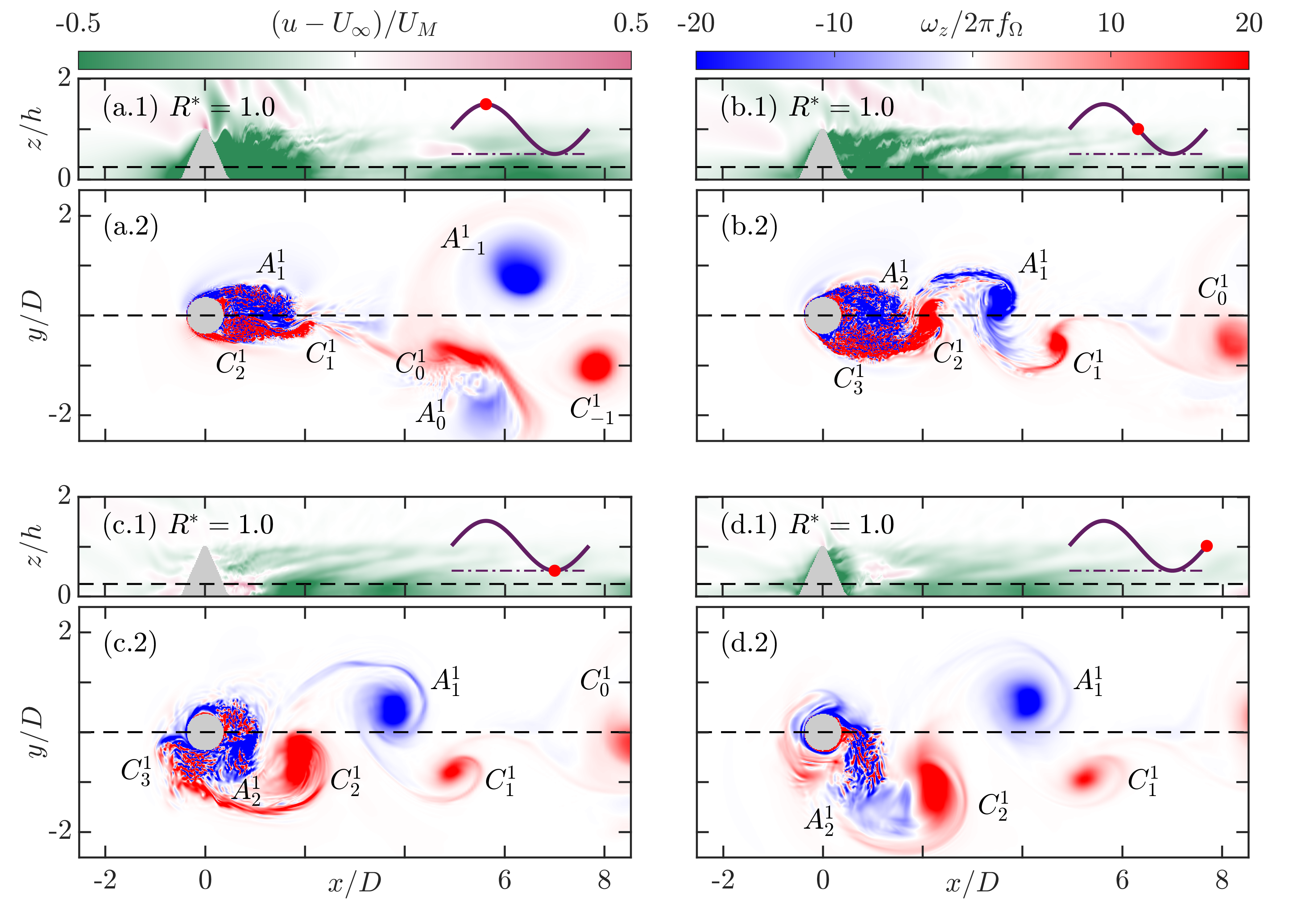}
    \caption{Evolution of the $\rs=1$ flow is illustrated by four different tidal phases $\phi$ in the 11th cycle: (a) $\pi/2$, (b) $\pi$, (c) $3\pi/\R{2}$ and (d) $2\pi$.
    At each tidal phase, the top and bottom sub-panels present: (\_.1) the streamwise velocity deviation on a vertical transect at $y/D=0$, and (\_.2) the vertical vorticity on a horizontal transect at $z/h=0.25$.
    The sinusoid plotted as insets in (\_.1) represents the variation of the free-stream velocity $\ui$.
    The red dot indicates the tidal phase and the dot-dashed line represents $\ui=0$.
    The dashed lines indicate $z/h=0.25$ in (\_.1), and $y/D=0$ in (\_.2).}
    \label{fig:f2}
\end{figure*}
%TC:endignore

\R{We begin with a discussion of the tidal phasing of the flow, wherein the streamwise velocity ($u$) on a vertical plane at $y/D=0$ and the vertical vorticity ($\omega_z$) on a horizontal plane at $z/h=0.25$ are investigated at four different phases from a representative tidal cycle (the 11th one is chosen).}
% Four different phases from a tidal cycle (the 11th one is chosen) are discussed using the streamwise velocity ($u$) on a vertical plane at $y/D=0$ and the vertical vorticity ($\omega_z$) on a horizontal plane at $z/h=0.25$.
There is cycle-to-cycle variability in the vortex patterns owing to both wake turbulence and a low-frequency modulation by the Coriolis force.
Nevertheless, there are overall trends in the vortex dynamics.
These trends will be compared among the cases with $\rs=\uc/\ut=1$, 0.5, and 0 (pure tide). The distribution of wake vortices is a function of the tidal phase. To facilitate the discussion, individual vortices in figures are labeled for each case: $A^{\rs}_{i}$ and $C^{\rs}_{i}$ correspond to an anticyclonic and a cyclonic vortex, respectively, of index $i$, from case $\rs$. For example, $A^{0.5}_{2}$ is an anticyclonic vortex for case $\rs = 0.5$ and index 2, and tracked as the phase progresses.

\subsection{Strength of the mean current equals the tidal amplitude ($\rs=1$)}
The discussion of temporal dynamics commences with the $\rs=1$ case, where the free-stream current is always eastward with $U_\infty \ge 0$. Insofar as wake eddies, \R{although} the value of $f^*=2.35$\R{,} there is sufficient time for a Strouhal pair to be shed from the topography between $\phi = \pi/2$ to $\pi$.
% \R{due to the influence of $\ut=\uc$}.
This pair of cyclonic and anticyclonic vortices is able to escape the near wake during most tidal cycles and propagate downstream as coherent wake eddies. However, the vortices formed during the latter half of the cycle, when the tidal component is westward (negative), interact with the near wake and often are not able to propagate out of the near wake as coherent wake eddies.

\begin{enumerate}
\item
The tidal phasing is elaborated starting with $\phi = \pi/2$ when the free-stream velocity ($\ui$) attains its maximum value of $\um=\uc+\ut$.
A near wake with a large velocity defect ($u<\ui$) is seen immediately downstream of the seamount ($0<x/D\le1$) in \fref{fig:f2}(a.1), below the lee waves emanating from the top of the seamount.
A thin hydraulic jet with a thickness $\sim \um/N$ is present at the interface between the lee waves and the velocity-defect region.
The horizontal slice of the vertical vorticity ($\omega_z$) in \fref{fig:f2}(a.2) reveals near-wake turbulence (multiscale vorticity).
Vortices shed earlier are also seen farther east, e.g., the anticyclonic vortex $A^1_{-1}$, whose signature is also present in the velocity defect as a zone of heightened negative $u - U_\infty$ in \fref{fig:f2}(a.1).
In fact, the cyclonic and anticyclonic vortices $C^1_{-1}$ and $A^1_{-1}$, respectively, form a Strouhal pair shed during the first half of the previous (10th) tidal cycle.

\item \R{Despite $f^*=2.35$ suggesting a shedding period of approximately half the tidal cycle, the relatively large $\ut=\uc$} results in the shedding of the near-wake vortices $C^1_1$ and $A^1_1$ of \fref{fig:f2}(a.2) as a well-formed Strouhal pair \R{within a quarter tidal cycle, as seen at $\phi=\pi$} in \fref{fig:f2}(b.2).
The cyclonic vortex $C^1_2$ is better developed, and a new anticyclonic vortex $A^1_2$ rolls up in the recirculation region.
Thin, inclined layers of enhanced velocity defect are more visible in \fref{fig:f2}(b.1) compared to \fref{fig:f2}(a.1), and a comparison with the vorticity field in \fref{fig:f2}(b.2) reveals their link to the peripheries of the vertically \R{coupled} vortices (see, for example, near $x/D=1$).
% Due to the proximity of the inclined vertically \R{coupled} vortices at $\phi=\pi$, the streamwise extent of the velocity-defect region appears longer in \fref{fig:f2}(b.1).
Signatures of vertical overturns are seen in these thin layers, and are associated with the vertical shear at the peripheries of the coherent vortices.
The velocity of the hydraulic jet above the recirculation zone is halved in \fref{fig:f2}(b.1) since the $\ui/N$ scale is halved from $\phi=\pi/2$ to $\phi=\pi$.
The wavelength of newly generated lee waves, although not very visible, is also shortened.

\item The current decelerates to $\ui =0$ at $\phi=3\pi/2$. The hydraulic jet is absent and the lee waves are weakened in \fref{fig:f2}(c.1).
Velocity defect, albeit weaker, is still seen in the vertical transect because of the vortices that were shed earlier.
The anticyclonic vortex $A^1_2$ in \fref{fig:f2}(b.2) is formed in the recirculation region during the deceleration of $\ui$, but it is not shed downstream and remains near the topography.
From $\phi=\pi$ to $3\pi/2$, it moves southward and interacts with the adjacent cyclonic vortex $C^1_2$ to form a vortex pair.
The formation of the vortex pair before $A^1_2$ fully escapes the near wake in turn results in the upstream (westward) advection of the turbulent near wake, despite $\ui$ remaining non-negative throughout the tidal cycle.
Note that a second cyclonic vortex $C^1_3$, carried over from the southern half of the recirculation region in \fref{fig:f2}(b.2), is also advected upstream (westward) of the seamount.
At phase $3\pi/2$, although $\ui = 0$, strong turbulence with fine-scale variability is seen in the near wake owing to the interactions between the vortical structures generated earlier and the topography.
As a result of these interactions, turbulence is also seen upstream of the seamount in \fref{fig:f2}(c.2).

\item As the current accelerates back to $\ui=\uc$ at phase $\phi=2\pi$, the vortex pairs $C^1_1$-$A^1_1$ and $C^1_2$-$A^1_2$ and the turbulent near wake seen at $\phi=3\pi/2$ are advected downstream (\fref{fig:f2}d.2), along with previously shed vortices farther east of the seamount.
Note that the $C^1_2$-$A^1_2$ vortex pair has moved away from $y/D=0$ because of a self-induced southward velocity.
The cyclonic vorticity associated with $C^1_3$ is not shed as a coherent vortex as it interacts with $A^1_2$ and breaks down into turbulence.
The streamwise velocity field in \fref{fig:f2}(d.1) also reflects the downstream advection of previously existing features, except immediately downstream of the seamount.
Remnants of upstream velocity defect seen in \fref{fig:f2}(c.1), albeit weak, are present at the end of the tidal cycle.
%Positive free-stream velocities from $\phi=3\pi/2$ to $2\pi$ have resulted in a short near wake region, with inclined layers of enhanced velocity defect.lee wave signatures are seen above the wake, along with a hydraulic jet between the two with an amplitude comparable to $\phi=\pi$ in \fref{fig:f2}(b.1), since the length scale $\ui/N=\uc/N$ at both instants.
\end{enumerate}

Stark differences exist between \fref{fig:f2}(b)~\&~(d) although both phases have $\ui = \uc$. The excursion number for the latter half of the tidal cycle is ${Ex^-(\rs=1)=1.63\sim10^0}$. The $\sim1$ value of $Ex^-$ for this case leads to the dissimilarity between $\phi = \pi$ and $\phi = 2\pi$ in \fref{fig:f2}, since the $C^1_2$-$A^1_2$ vortex pair, for example, has not been advected far away from the topography by $\phi=2\pi$.
Finally, a comparison between \fref{fig:f2}(d)~\&~(a) allows us to complete the tidal cycle.
In fact, the nonlinear evolution of the $C^1_2$-$A^1_2$ vortex pair, together with the turbulent breakdown of $C^1_3$, gives rise to vortical structures in cycle 12 that are analogous to the $C^1_0$-$A^1_0$ pair of cycle 11 in \fref{fig:f2}(a).

In summary, while there is some cycle-to-cycle variability, as revealed by Supplemental Fig.~S1, overall trends can be inferred.
During the first half of the tidal cycle for the $\rs=1$ case with $f^*=2.35$, when the tidal component is in the direction of the mean current, a Strouhal pair is shed.
Within the Strouhal pair, the second vortex tends to be larger and stronger (also see Supplemental Fig.~S1) because the generation of the first is hindered by the near wake dynamics of the previous cycle.
During the latter half of the tidal cycle, at least two new vortices are generated, but they are not shed as coherent vortices owing to low $\ui$.
Instead, they form a vortex pair close to the seamount and influence the dynamics of the near wake, even when $\ui=0$ at peak westward tidal flow.
The vortex pair is laterally displaced from the centerline ($y=0$) and then advected away from the seamount as $\ui$ grows stronger eastward.

%TC:ignore
\begin{figure*}[!t]
    \centering
    \includegraphics[scale=0.9]{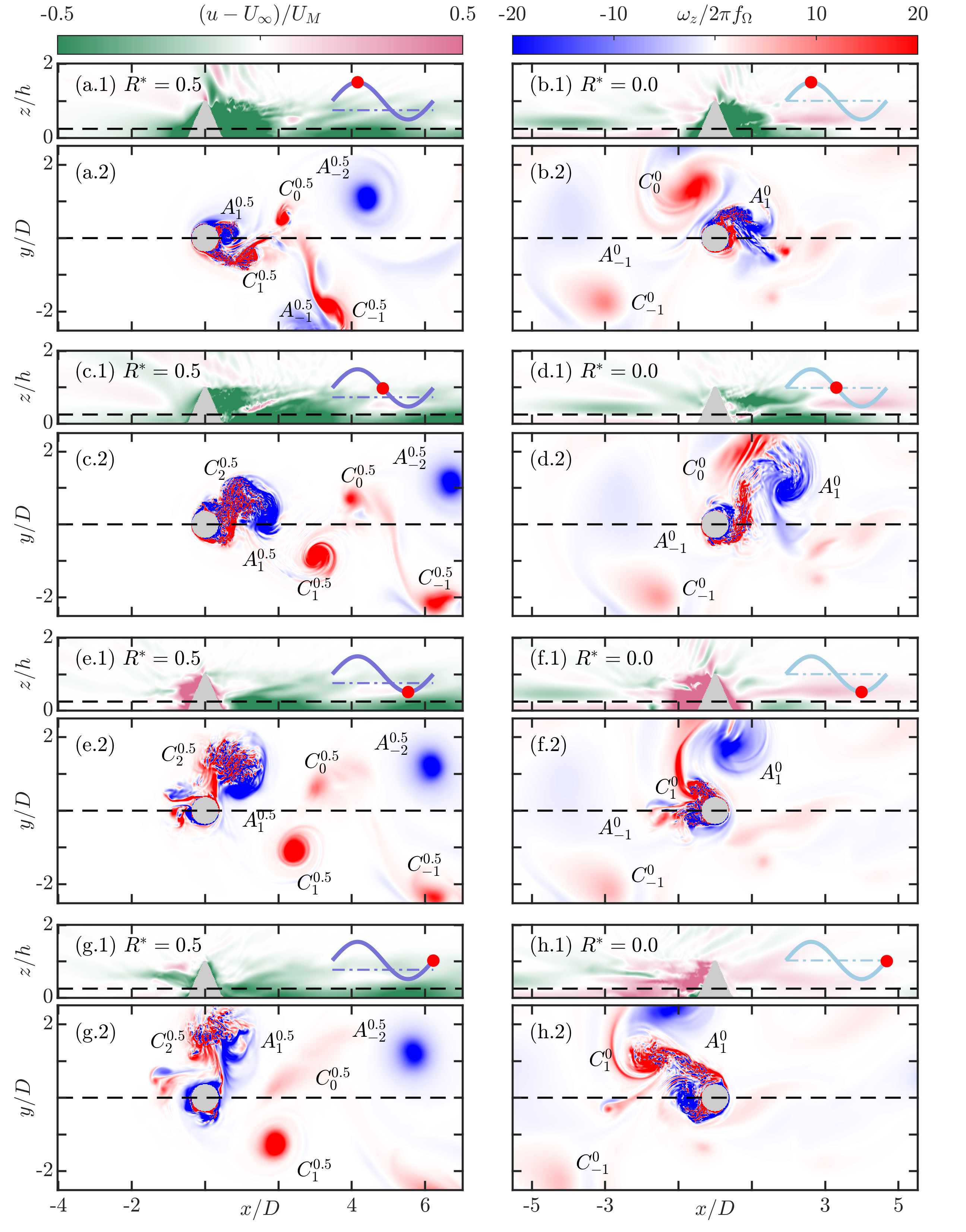}
    \caption{Same as \fref{fig:f2}, but for $\rs=0.50$ (left column) and $\rs=0.00$ (right column).
    Note the vertical shift in the $\ui=0$ indicator (dot-dashed line) with respect to the sinusoid in the insets, as $\rs$ is varied.
    }
    \label{fig:f3}
\end{figure*}
%TC:endignore

\subsection{Strength of the mean current equals half the tidal amplitude ($\rs=0.5$), and no mean current/pure tide ($\rs=0$)}
Moving to cases with $\rs = 0.5$ and $0$, qualitative differences in the wake eddies are observed.
Unlike $\rs=1$, both cases experience a westward current ($\ui<0$) during parts of the tidal cycle, i.e., when the sinusoids plotted in the insets of \fref{fig:f3}(a.1) and (b.1) lie below the respective dot-dashed lines.
Regardless, at $\phi=\pi/2$, eastward $\ui$ peaks at $\um=\uc+\ut$ for all $\rs$.

In the $\rs=0.5$ case, the $\uc$-based vortex shedding \R{frequency is half} that at $\rs=1$ --- see Table~\ref{tab:params}: $f^*=1.16$ and $2.35$, for $\rs=0.5$ and $1$, respectively.
Because of the slower vortex shedding, unlike $\rs=1$, two fully formed vortex pairs are unable to escape the near wake as coherent wake eddies during each tidal cycle.
Instead, at $\rs = 0.5$, a single vortex of alternating polarity and a Strouhal pair is shed eastward every tidal cycle.
Other vortices resulting from flow separation are observed, but do not remain coherent as they are advected away from the topography.
For $\rs=0$, on the other hand, $\uc = 0$, and hence the mean-current-based vortex shedding is irrelevant.
Instead, phase eddies are formed during tidal acceleration and shed during deceleration --- see observations of \citet{black1987eddy}, \citet{pawlak2003evolution} and \citet{wynne2022measurements}, for example --- instead of Strouhal pairs.

\begin{enumerate}
\item Similar to $\rs=1$, recirculation regions with large velocity defects are present for both $\rs=0.5$ and $0$, in \fref{fig:f3}(a.1) and (b.1), respectively.
Hydraulic jets with $\ui/N$ length scale and lee waves are also seen above the recirculation regions.
The near-wake $\omega_z$ fields in \fref{fig:f3}(a.2) and (b.2) correspond to different stages of vortex shedding.
Farther east, \fref{fig:f3}(a.2) reveals a previously shed vortex $A^{0.5}_{-2}$ and a $C^{0.5}_{-1}$-$A^{0.5}_{-1}$ vortex pair ($y/D\approx-2$) that were advected eastward away from the seamount.
A weaker cyclonic vortex $C^{0.5}_{0}$ is also seen, but it does not remain coherent until the end of the tidal cycle.

For $\rs=\uc/\ut=0$, the absence of mean advection implies that previously shed vortices remain in the vicinity of the seamount and potentially interact with the near wake.
This is evidenced in \fref{fig:f3}(b.2), where a few vortices ($C^0_0, A^0_{-1}, C^0_{-1}$) that were previously shed westward, and subsequently advected back towards the seamount during $0<\phi\le\pi/2$, are seen.
As will be seen, the northern vortex $C^0_0$ has a strong influence on the subsequent evolution of the wake vortices.

\item At the later phase $\phi=\pi$, \fref{fig:f3}(d.2) reveals that $C^0_0$ pairs with a newly formed (during $0 < \phi < \pi/2$) $A_1^0$, resulting in the shedding of a vortex pair $C^0_0$-$A^0_1$ to the north-east of the seamount.
A vortex pair $A^{0.5}_{1}$-$C^{0.5}_{2}$ is also seen in \fref{fig:f3}(c.2) for $\rs=0.5$, but the mechanism here is similar to $\rs=1$, wherein the cyclonic vorticity in the vortex pair originates from the southern half of the recirculation region, as opposed to a previously shed vortex for $\rs=0$.
Inclined layers of enhanced velocity defect are more clearly seen during the deceleration stage for both $\rs$, in \fref{fig:f3}(c.1) and (d.1), as was the case for $\rs=1$.
%The free-stream velocity $\ui=0$ at phase $\pi$ for $\rs=0$ due to the absence of the steady mean current, and as a result, lee waves and the hydraulic jet are absent in \fref{fig:f3}(d.1).

\item Later, at tidal phase $\phi=3\pi/2$, the current with $\ui<0$ attains a westward maximum for all $\rs<1$, and a near wake region with a `velocity defect' is seen to the west of the seamount in \fref{fig:f3}(e.1) and (f.1).
Lee wave signatures, along with the hydraulic jet, with length scales $\sim|\ui|/N$, are also seen west of the seamount, above the recirculation region.
As seen in \fref{fig:f3}(e.2) and (f.2), the vortex pairs shed earlier to the northeast are advected westward by the negative $\ui$ and northward by their self-induced velocities.
New recirculation regions are formed to the west of the seamount since $\ui$ is negative, and the near wake is better developed for $\rs=0$ because of a larger $\left|\ui\right|$, and a longer exposure to $\ui<0$.

The polarity of the northern and southern vortices within the recirculation region is reversed for a westward wake compared to an eastward wake --- for example, the northern vortex $C^0_1$ in the recirculation region is cyclonic in \fref{fig:f3}(f.2), but $A^0_1$ is anticyclonic in \fref{fig:f3}(b.2).
An implication of the polarity reversal, as the current gradually changes its direction, is the shedding of residual vorticity from the recirculation region on the eastern side as small vortices and/or vortex pairs, as seen immediately towards the west of the seamount in \fref{fig:f3}(e.2) and (f.2).
An example is the small vortex pair near $(x/D,y/D=2,0.5)$ in \fref{fig:f3}(b.2).

\item At the end of the tidal cycle, the flow features seen in \fref{fig:f3}(g.2) are reminiscent of \fref{fig:f3}(e.2), albeit the turbulent breakdown of the northern $A^{0.5}_{1}$-$C^{0.5}_{2}$ vortex pair. This is reasonable for $\rs=0.5$, given its low $Ex^-=-0.61$, i.e., small westward advection by the current.
In contrast, $\rs=0$ has an order of magnitude larger $Ex^-=-2.86$.
Therefore, \fref{fig:f3}(h) is almost a mirror image of \fref{fig:f3}(d), with a vortex pair $A^{0}_{1}$-$C^{0}_{1}$ shed to the northwest.
%Note that the fields presented in \fref{fig:f3}(g) correspond to $\phi$ close to, but less than $2\pi$ (red marker in the inset centered marginally below the dot-dashed $\ui=0$ line).

The vertical sections of defect velocity have distinctive characteristics.
For case $\rs = 0.5$, \fref{fig:f3}(g.1) shows the persistence of large velocity defects west of the seamount at the end of the tidal cycle, despite $\rs=0.5$ experiencing westward free-stream velocities for only a fraction of the tidal cycle. For case $\rs = 0$, there are well-formed wakes at $\phi = \pi/2$ (\fref{fig:f3}b.1) and $\phi = 3\pi/2$ (\fref{fig:f3}f.1) on the eastward and westward sides, respectively, with similar geometry but of opposite sign. Although the free-stream current is zero at $\phi = \pi$ (\fref{fig:f3}d.1) and $2\pi$ (\fref{fig:f3}h.1), there is significant local velocity near the topography.
\end{enumerate}

In summary, over the course of the 11th tidal cycle, a cyclonic vortex and a vortex pair were shed downstream, to the south and north of the $y/D=0$ line, respectively, for the $\rs=0.5$ case ($f^*=1.16$).
Similar to $\rs=1$, the coherent vortex was shed during the eastward tide, and the vortex pair was generated when the tide flowed westward.
\R{For $\rs=0.5$,} the \R{subsequent (12th) tidal} cycle (figures not shown) shows mirrored (around the $y=0$ line) vortex shedding: an anticyclonic vortex shed to the north of $y/D=0$ and a vortex pair to the south.

%TC:ignore
\begin{figure*}[!t]
    \centering
    \includegraphics[scale=0.714]{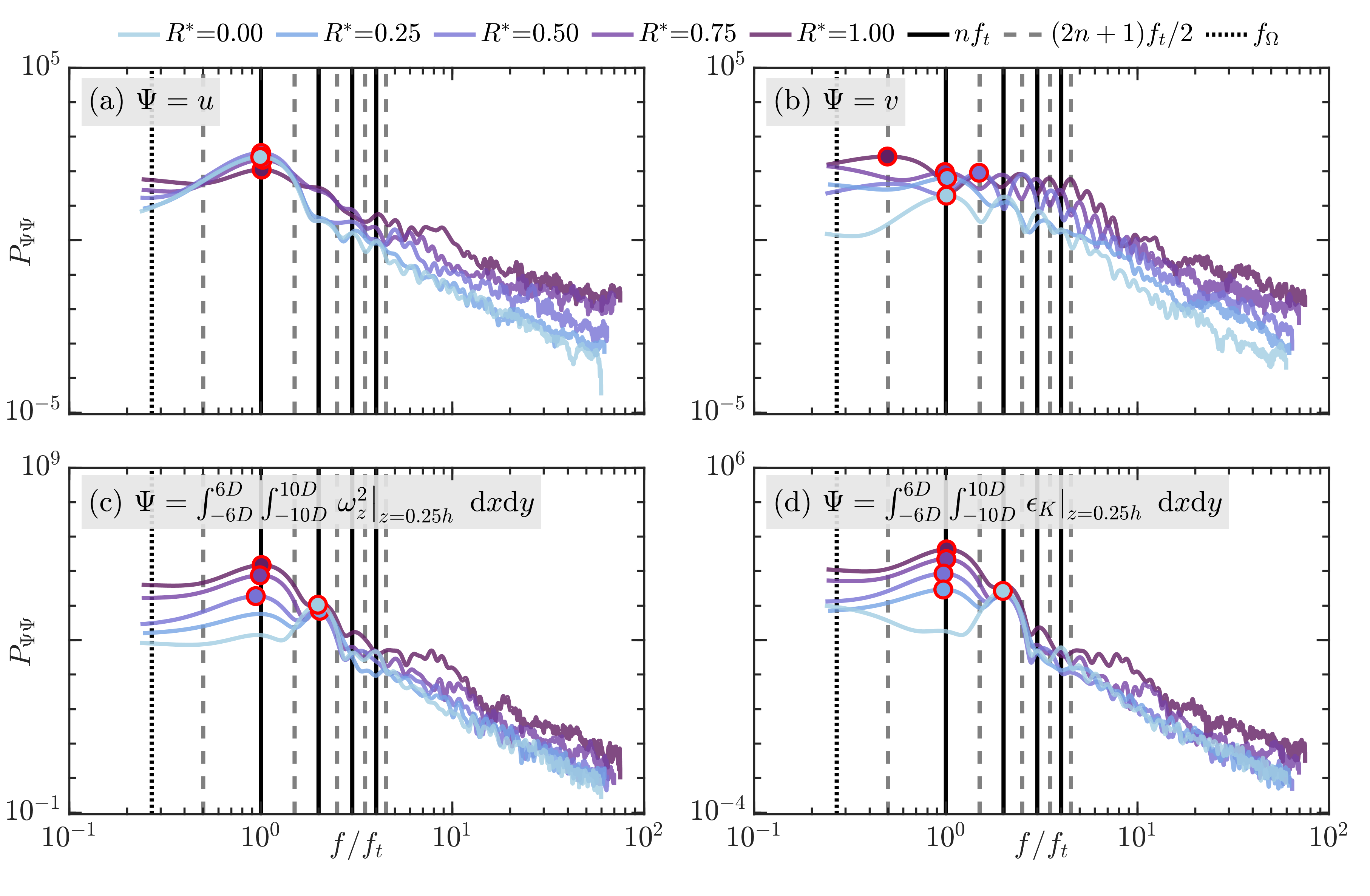}
    \caption{Temporal spectra of (a) the streamwise velocity $u$, (b) the spanwise velocity $v$, (c) squared vertical vorticity $\omega_z^2$, and (d) dissipation rate of kinetic energy per unit mass $\ek$, for the five $\rs$ considered in this study.
    In (a) \& (b), the spectra are calculated from measurements at a point $(x/D,y/D,z/h)=(1.5,0,0.25)$.
    The spectra in (c) \& (d) are for the horizontal-area integrals of the respective quantities, at $z/h=0.25$.
    The vertical lines indicate the background rotation $f_\Omega$ (dotted), tidal frequency \& its harmonics $nf_t$ (black solid), and half the tidal frequency \& its odd multiples $(2n+1)f_t/2$ (gray dashed), with $n=0,1,2,3,4$.
    Red circles indicate the most dominant peak in each spectrum.}
    \label{fig:f4}
\end{figure*}
%TC:endignore

For the pure tide ($\rs=0$), however, a potentially bistable state was seen.
Anticyclonic and cyclonic vortices were shed to the north of $y/D=0$, during the first (eastward tide) and second (westward tide) halves of the tidal cycle, respectively.
The newly shed vortices combined with previously shed vortices to form vortex pairs that advected away in the $y$ direction due to self-induced velocities.
During either half of the cycle, the presence of previously shed vortices north of the seamount influenced the recirculation region, locking the vortex shedding to the north of $y/D=0$ \R{across multiple tidal cycles}.
The sustained northward vortex pair shedding is likely one of two possible scenarios --- we believe the system is equally likely to lock onto a vortex-pair shedding towards the south of $y/D=0$.

\subsection{Temporal spectra for $\rs\in\{1.00,0.75,0.50,0.25,0.00\}$}
The tidal phasing of the wake vortex dynamics demonstrates that wake dynamics vary considerably with $R^*$, in tidally dominated flows past 3D topography.
However, comprehensively mapping these complex wake dynamics in the ocean can be challenging.
Localized velocity measurements are obtained from moorings and ADCPs during field experiments, and a spectral analysis can reveal links to the dominant forcing in the system (i.e., the tide).
Hence, we present the temporal spectra of $u$ and $v$ measured at a point inside the near wake for all $\rs$ in \fref{fig:f4}(a) and (b), respectively.
To link these velocity spectra to the vortex dynamics in observed in the simulations, temporal spectra of area-averaged vertical vorticity are presented in \fref{fig:f4}(c).
Since dissipation/mixing is a primary focus of this work, the power spectral density of the area-integrated $\ek$ is presented in \fref{fig:f4}(d).

Each temporal spectrum in \fref{fig:f4} was calculated from a data set of around 2700 samples collected over 18 tidal cycles, after subtracting the temporal mean.
The sampling frequency is almost 5 minutes so that the highest frequency resolved in the post-processing is $75f_t$ (Nyquist criterion), where $f_t=2.237\times10^{-5}$~s$^{-1}$ is the $M_2$ tidal frequency.
% 5 min sampling (sampling freq = 1/(5*60)) means the highest frequency resolved is (1/2)(1/(5*60)) = 1.67e-3 Hz = 74.5 f_t; f_t = 1/( (24+50/60)/2 * 3600) = 2.237e-5 Hz
Welch's method was used to obtain spectra, and each data set was divided into nine segments with an overlap of 10\%, with a Hanning window to reduce edge effects.
% As a result, the limit of frequencies that can be resolved in \fref{fig:f4} is approximately $f_t/2$.
Recall, that the Coriolis parameter ($2\pi f_\Omega$) at $15^\circ$~N considered in the simulations corresponds to $f_\Omega\approx f_t/4$ (dotted line in \fref{fig:f4}).

The temporal spectra of the streamwise ($u$) and spanwise ($v$) velocity are measured at the point $(x/D,y/D,z/h)=(1.5,0,0.25)$ in the near wake to compare with the previous work on tidally modulated ($\rs > 1$) flow \citep{puthan2022wake}.
The $u$-spectra in \fref{fig:f4}(a), for all five $\rs$, present a dominant peak at $f_t$, the streamwise forcing frequency in the system.
This is in agreement with the observations for tidally modulated flows past a seamount in \citet{puthan2021tidal}.
Spectral peaks are also seen at integer multiples of $f_t$.

The $v$-spectra at the near-wake measurement location {\em do not all} present a dominant peak at $f_t$ in \fref{fig:f4}(b)\R{, as a result of tidal synchronization of vortex shedding~\citep{puthan2021tidal}. This means that the} intrinsic vortex shedding frequency $f_{vs,c}$ of the mean current matters to the $v$-spectra.
The peak for $\rs=1$ is at the subharmonic $f_t/2$ and its integer multiples of $f_t/2$, similar to the $\rs =1$ case of \citet{puthan2022wake} that overlaps with the present strong-tide regime.
For the next smaller $\rs$ of $0.75$ and $0.5$, the peaks are $f_t$ and $3f_t/2$, respectively.
The $\rs=0.5$ case exhibits an additional local peak at $f_t/2$.
For the two smallest $\rs = 0.25$ and $\rs = 0$ with the weakest mean current ($\uc < 0.5\ut$), the dominant peak is at $f_t$, similar to the $u$-spectra.
In all cases, spectral peaks are seen at integer multiples of the dominant frequency.

The coherent vortex dynamics near the seamount is reflected in the power spectra of the area-integrated $\omega_z^2$ presented in~\fref{fig:f4}(c).
The area integral is performed on the horizontal transects at $z/h=0.25$, presented earlier in \fref{fig:f2} and \ref{fig:f3}).
A dominant peak is seen at the forcing frequency $f_t$ for the three cases with strong mean currents $\rs=1$, $0.75$, and $0.5$.
Although the two cases with the weakest mean currents, $\rs=0.25$ and $0$, also show peaks at $f_t$, the frequency $2 f_t$ is more dominant.
This is because for both $\rs$, vortices (and/or vortex pairs) are shed during the latter half of the tidal cycle as well, as suggested by the relatively large negative magnitudes of excursion number $Ex^-=-1.74$ and $-2.86$, for $\rs=0.25$ and $0$, respectively.

The spectra for $\rs=1$, $0.75$ and $0.5$ are qualitatively similar to corresponding $\omega^2_{z}$ spectra in \fref{fig:f4}(c).
This, however, does not mean that the horizontal shear is the primary source of dissipation.
As will be demonstrated later, a significant portion of the wake dissipation occurs by vertical shear at the peripheries of the \R{vertically coupled} coherent horizontal vortices, and this link to $\omega_z$ is the reason for the qualitative similarity between the $\ek$ and $\omega^2_{z}$ spectra and their peaks at $f_t$.
The dominant peak of the $\ek$-spectrum for $\rs=0.25$ is at $f_t$ with another slightly smaller peak at $2f_t$ (the peak in its $\omega^2_{z}$ spectrum).
The dominant peak of the $\ek$-spectrum for $\rs=0$ is clearly at $2f_t$, matching its dominant peak in the $\omega^2_{z}$ spectrum.

%TC:ignore
\begin{figure*}[!t]
    \centering
    \includegraphics[scale=0.714]{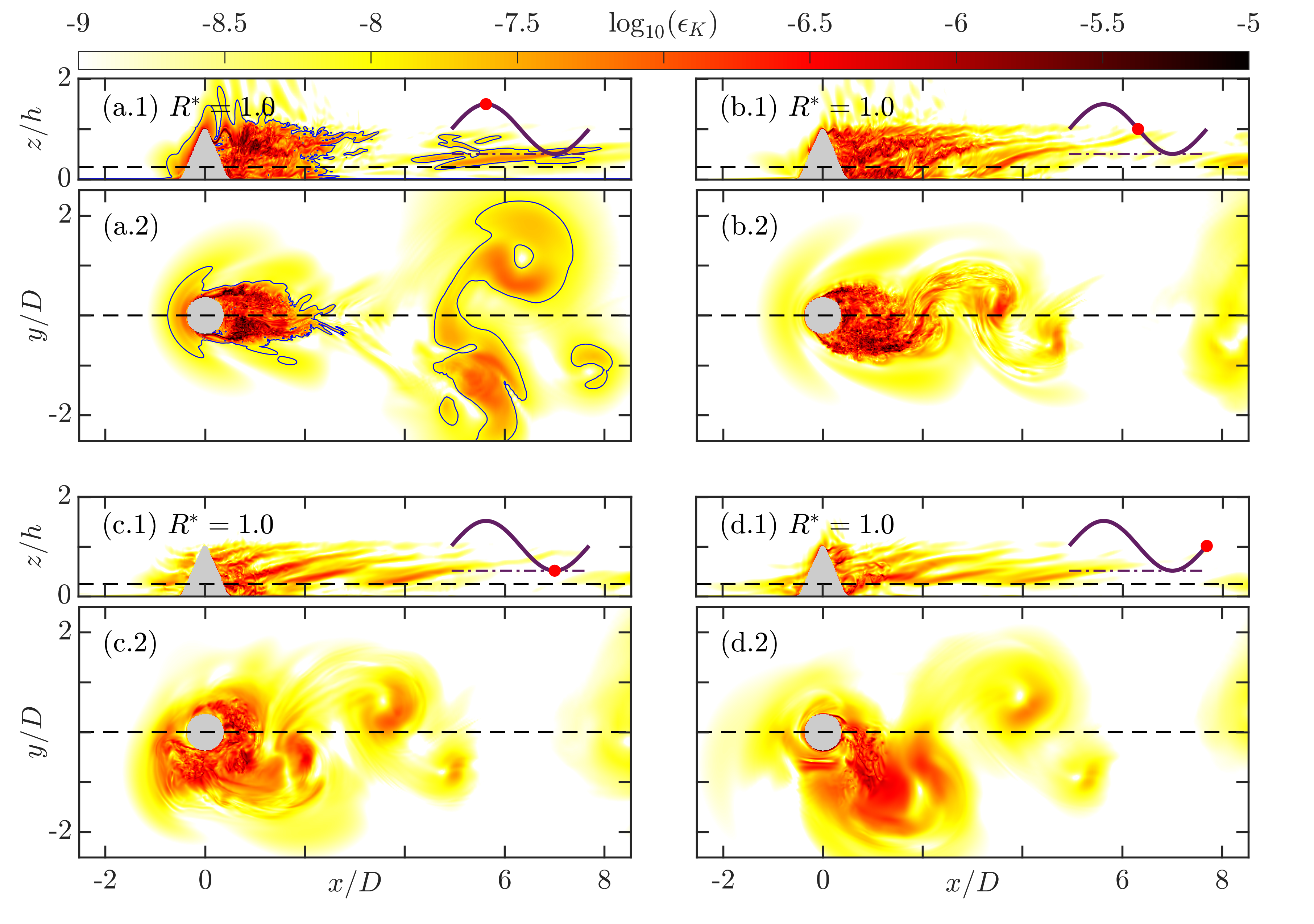}
    \caption{Dissipation rates $\ek$ (in~\w) for $\rs=1$, at four different tidal phases: (a) $\pi/2$, (b) $\pi$, (c) $3\pi/\R{2}$ and (d) $2\pi$.
    At each tidal phase, the top and bottom sub-panels correspond to a vertical transect at $y/D=0$ and a horizontal transect at $z/h=0.25$, respectively.
    The dashed lines in the vertical and horizontal transects indicate $z/h=0.25$ and $y/D=0$, respectively.
    The $\ek$ fields are to be compared with the velocity and vorticity fields of \fref{fig:f2}.
    The blue contour in panel (a) corresponds to a dissipation rate $\ek=10^{-8}$~\w.}
    \label{fig:f5}
\end{figure*}
%TC:endignore

\section{Instantaneous dissipation rates of kinetic energy}
\label{ssec:instantaneous}
As a first step towards understanding mixing in such tidally dominated topographic flows, the tidal phasing of the dissipation rate of kinetic energy per unit mass ($\ek$) is investigated.
%Figure~\ref{fig:f5} presents the spatial distribution of $\ek$ at the four tidal phases presented in \fref{fig:f2}, for $\rs=1$.
It will be shown that there is substantial phase-dependent dissipation (\esim{-5}) in the near wake ($x/D\le2.5$).
Patches of dissipation \esim{-7}, associated with wake eddy instabilities, are seen further downstream up to $x/D\approx6$ and laterally up to $y/D\approx\pm2$.

%TC:ignore
\begin{figure*}[!t]
    \centering
    \includegraphics[scale=0.714]{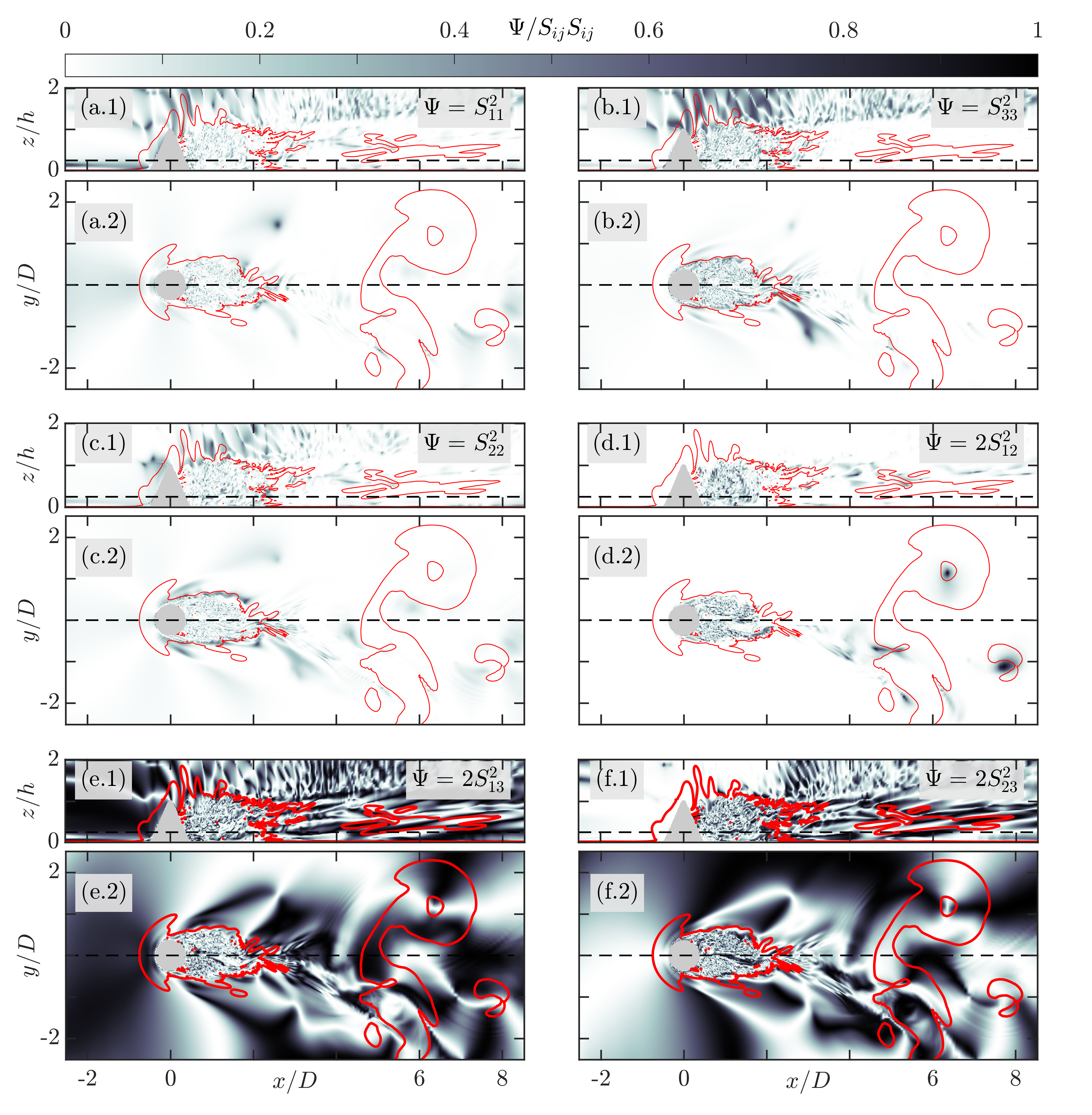}
    \caption{Contribution from the six unique components of the strain-rate tensor (a) $S_{11}$, (b) $S_{33}$, (c) $S_{22}$, (d) $S_{12}$, (e) $S_{13}$, and (f) $S_{23}$, to the dissipation rate $\ek$ at tidal phase $\phi=\pi/2$ for $\rs=1$, which was presented in \fref{fig:f5}(a).
    The red contour in each panel corresponds to $\ek=10^{-8}$~\w, and matches the blue contour in \fref{fig:f5}(a).}
    \label{fig:f6}
\end{figure*}
%TC:endignore

\begin{enumerate}
\item At $\phi=\pi/2$ (\fref{fig:f5}a), the near wake displays large dissipation rates with \esim{-5} as the strongest values.
The vertical transect in \fref{fig:f5}(a.1) reveals that the hydraulic jet, which breaks down into turbulence approximately one wavelength downstream of the seamount, is a site of intense dissipation.
The recirculation region below the jet displays significant $\ek$ as well, albeit an order of magnitude smaller.
Non-trivial dissipation is also seen near the site of lee wave generation, but with much smaller values, \esim{-8}.
The vortices shed earlier and advected eastward to $x/D\approx6$ also display \esim{-7} along their peripheries in \fref{fig:f5}(a.2).
Thus, instability of wake eddies results in significant dissipation at a large downstream distance of $x/D\approx6$.
The strongest dissipation rates (\esim{-5}) in the horizontal transect are associated with the attached shear layers which form when, under strong stratification, the flow goes around the 3D topography and separates laterally.

\item The `elongated' near wake seen in \fref{fig:f5}(b.1) at $\phi=\pi$, owing to the wake-vortex dynamics presented in \fref{fig:f2}(b), exhibits large \esim{-6}.
A comparison with \fref{fig:f2}(b.1) reveals that the inclined layers of enhanced velocity defect that show vertical overturns are sites of strong dissipation.
The link between the inclined layers to vortex peripheries is further supported by correspondence between the \esim{-7} signatures at $x/D=3.5$ in \fref{fig:f5}(b.1) and (b.2).
In general, the near-wake dissipation regions are more spread out at $\phi=\pi$ ($\ui=\uc$) compared to $\phi=\pi/2$ ($\ui=\uc+\ut$).
However, the strongest dissipative features, i.e., the hydraulic jet and the attached shear layers, do not contribute nearly as much to wake dissipation at $\phi=\pi$ as they do at $\phi=\pi/2$.
Hence, as will be shown, the volume-integrated value of $\ek$ at $\phi=\pi$ is lower than at $\phi=\pi/2$.

\item The inclined layers of \esim{-7} persist when $\ui =0$ at tidal phase $\phi=3\pi/2$ in \fref{fig:f5}(c.1).
They are associated with the peripheries of previously shed vortices seen in \fref{fig:f5}(c.2).
The turbulent recirculation region in \fref{fig:f5}(c.2), which almost surrounds the topography, shows \esim{-7}.
An interesting finding, as a result, is the presence of `rising dissipation signatures' {\em upstream} of the seamount, despite $\ui$ remaining non-negative throughout the tidal cycle.
%lee waves and the hydraulic jet are absent since $\ui=0$, but one could argue the presence of a dissipative attached shear layer in \fref{fig:f5}(c.2) at the northern boundary of the recirculation region, and to the southwest of the seamount (immediately south of the $y/D=0$ line).

\item Ultimately, at the end of the tidal cycle in \fref{fig:f5}(d), the dissipative structures seen earlier are advected downstream, as the current accelerates back up to $\ui=\uc$.
The attached shear layers at the boundary of the small, newly-formed recirculation region present enhanced \esim{-6} (c.f.~\fref{fig:f2}(d.2) and \fref{fig:f5}(d.2)).
The vertical transect in \fref{fig:f5}(d.1) shows a dissipative hydraulic jet, and lee waves with non-trivial \esim{-8}.
Signatures of nascent vertical overturns are also seen, at $(x/D,z/h=1,1)$, for example, with \esim{-7}.
The dissipative structures seen upstream of the seamount in \fref{fig:f5}(c) persist, albeit with weaker $\ek$ magnitudes.
\end{enumerate}

\R{To understand the role played by the subgrid-scale model in the dissipation rates observed in our LES, the ratio of subgrid-scale viscosity and molecular viscosity ($\nu_{SGS}/\nu$) for $\rs =1$ is presented in Supplemental Fig.~S2.
As indicated by the probability density functions of $\nu_{SGS}/\nu$, the median value of the ratio is vanishingly small ($\sim10^{-14}$), and non-trivial $\nu_{SGS}$ is localized to the near wake.
Thus, by design, the resolution of the turbulence by the chosen grid is very good.}

%TC:ignore
\begin{figure*}[!t]
    \centering
    \includegraphics[scale=0.9]{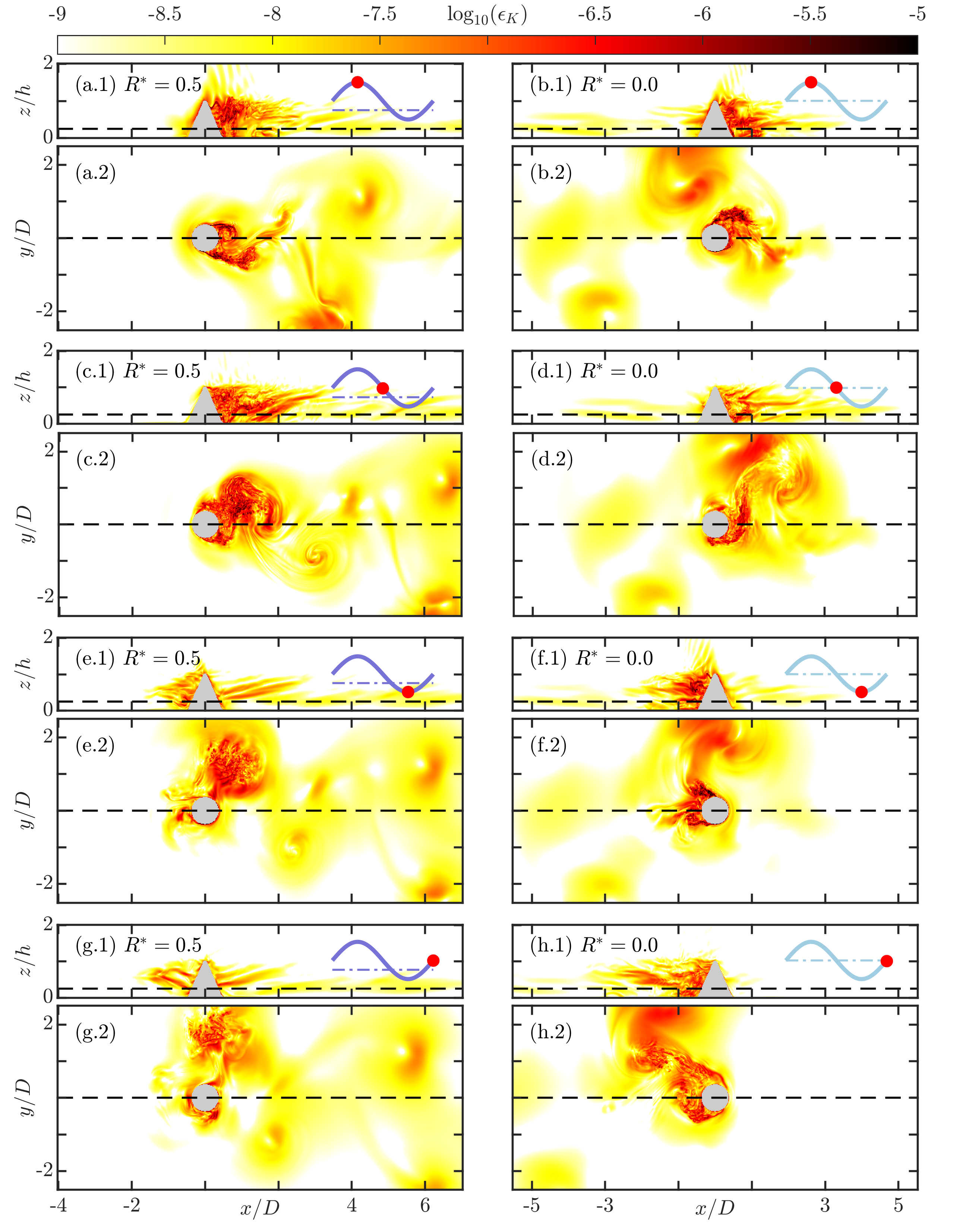}
    \caption{Same as \fref{fig:f5}, but for $\rs=0.50$ (left column) and $\rs=0.00$ (right column). The dissipation rate fields are to be compared with the velocity and vorticity fields of \fref{fig:f3}.}
    \label{fig:f7}
\end{figure*}
%TC:endignore

To gain some insight into the underlying mechanisms for the dissipation seen in the system, contributions from each of the six independent components of the \R{instantaneous} strain rate tensor to the dissipation rate $\ek$ presented in \fref{fig:f5}(a) are presented in \fref{fig:f6}.
The blue contours in \fref{fig:f5}(a), corresponding to a dissipation rate of ${\ek=10^{-8}}$~W~kg$^{-1}$, are also plotted in red in all 6 panels of \fref{fig:f6} to delineate regions of strong dissipation.

Outside the turbulent near wake, the normal strain rate $S_{11}$ in \fref{fig:f6}(a) contributes primarily to the dissipation associated with the lee waves above the seamount.
Contributions to the wake waves, e.g. \citep{gola2024internal}, and region of `upstream-blocking'~\citep{winters2014topographic} are also seen, but these regions correspond to low $\ek$ in \fref{fig:f5}(a).
The vertical strain-rate $S_{33}$ in \fref{fig:f6}(b) dominates the lee wave dissipation rates, and also contributes to the wake waves that are eastward and above the seamount wake.
%The spanwise normal strain-rate, $S_{22}$ in \fref{fig:f6}(c), plays a smaller role outside the near wake, with its contribution limited to regions with nonlinear interaction between waves.

%TC:ignore
\begin{figure*}[!t]
    \centering
    \includegraphics[scale=0.714]{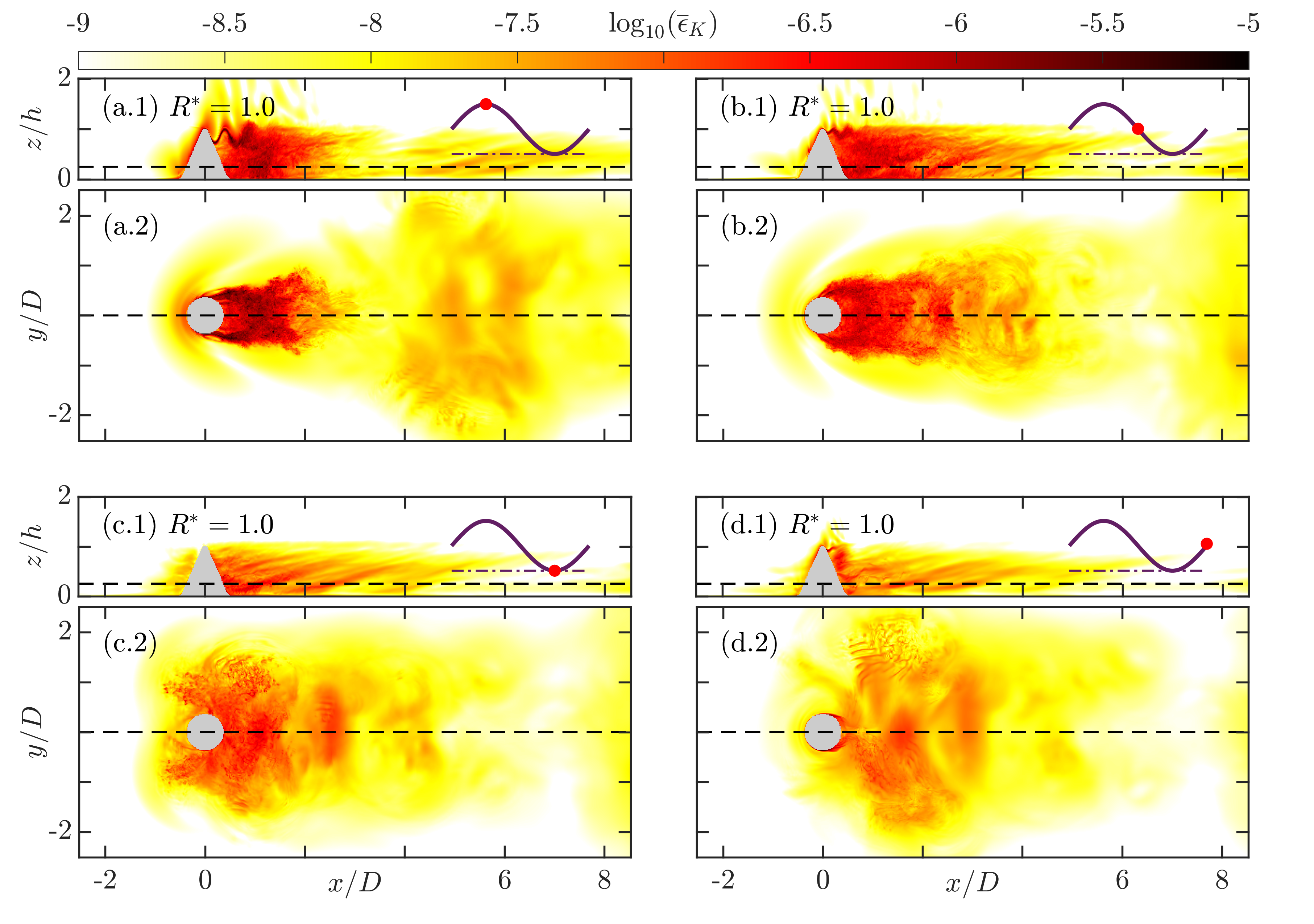}
    \caption{Same as \fref{fig:f5}, but the quantity presented is the phase-averaged dissipation rate $\oek$, in W~kg$^{-1}$.}
    \label{fig:f8}
\end{figure*}
%TC:endignore

The horizontal shear of horizontal velocities $S_{12}$ does not play a role downstream of the near wake in the vertical plane, in \fref{fig:f6}(d.1).
Its contribution is maximized at the centers of shed vortices (c.f. \fref{fig:f6}(d.2) and \fref{fig:f2}(a.2)), but these are not regions of strong dissipation (c.f. \fref{fig:f6}(d.2) and \fref{fig:f5}(a.2)).
As a result, even though vertically \R{coupled} horizontally eddying motions dominate the wake, the seen dissipation rates $\ek$ are not because of the horizontal strain $S_{12}$.
Instead, the two vertical shear components $S_{13}$ and $S_{23}$ dominate, especially in the inclined layers of strong dissipation at the peripheries of the coherent vortices.
Note, that $S_{23}$ contributes to the dissipation associated with the wake waves, but not the lee waves.

Within the immediate near wake, the turbulent flow is closer to three-dimensional, and the contributions from the different strain-rate components are of the same order of magnitude.
The dissipation in the attached shear layers, before they transition to turbulence, is primarily due to the horizontal shear $S_{12}$, in addition to $S_{13}$.
Similarly, the laminar dissipation in the hydraulic jet, prior to transition, is due to $S_{11}$, $S_{33}$, and $S_{13}$.

For completeness, \fref{fig:f7} presents $\ek$ for the cases with $\rs=0.5$ and 0 (pure tide) whose velocity/vorticity fields were presented in \fref{fig:f3}. For brevity, we refrain from an elaborate description. Similar to $\rs =1$, strongest \esim{-5} is associated with hydraulic jets and attached shear layers with additional contributions from inclined velocity defect layers in the near wake and peripheries of wake eddies. The phase/spatial distribution of $\ek$ corresponds to that of the velocity and vorticity fields shown in \fref{fig:f3}.
Both $\rs=0.5$ and $0$ show strong dissipation signatures west of the seamount by the end of the tidal cycle.
These `upstream' signatures are stronger for smaller $\rs$ because the more dominant the tide, the stronger (and longer) the westward $\ui$ experienced by the topography.

\section{Phase-averaged dissipation and mixing}
\label{ssec:phase-averaged}

This section is devoted to the dependence of dissipation and mixing statistics on the relative strength of the tidal component, $\rs$. A tidal phase average is employed to obtain a robust statistical measure. The volume integrated dissipation --- its phase dependence, and cycle average --- varies systematically with $\rs$. These wake dissipation values owing to the `stirring rod' turbulence of 3D topographies are compared with the internal wave flux (wave dissipation) and found to be substantially larger. The mixing of the buoyancy field by turbulence is also quantified.

The wake structures and corresponding dissipation signatures discussed in \S~\ref{ssec:dynamics} and \S~\ref{ssec:instantaneous} constitute one realization of an ensemble of plausible scenarios arising from vortex-vortex and vortex-topography interactions subject to turbulence. In addition, there is low-frequency modulation due to background rotation.
Supplemental Fig.~S1 presents ten realizations of the $\phi=\pi/2$ tidal phase for $\rs=1$, taken across ten consecutive tidal periods. The ensemble exhibits fluctuations around a well-defined state: two developed wake eddies in $2< x/D< 6$ preceded by eddy formation at the topography, $0 < x< D < 2$.
Phase-averaged statistics, obtained here by utilizing data from ten consecutive cycles from the tenth cycle ($9 \le tf_t < 19$), will be used to characterize the flow statistics.

Figure~\ref{fig:f8} presents the phase-averaged dissipation, denoted by $\oek$, for $\rs=1$ at four different tidal phases, with \fref{fig:f5} showing one of the ten realizations used for the phase averaging.
The $\oek$ fields at all four $\phi$ are relatively smooth compared to the instantaneous $\ek$ fields, and approximately symmetric about the $y/D=0$ line (except at $\phi = 2\pi$, to be discussed shortly), indicating that the 10-cycle phase average is adequate.
\R{The north-south symmetry of $\oek$ is also in agreement with the cyclonic-anticyclonic symmetry in a vertical shear dominated regime, as discussed by \citet{srinivasan2021high}.}
Importantly, as summarized below, the large $\ek$ values encountered during the 11th cycle are representative of the phase average. The hot spots of dissipation, although somewhat spatially smeared, are robust to phase averaging. For example, at $\phi=\pi/2$, the $\oek$ fields present the strongest dissipation rates \oesim{-5} at the hydraulic jet and the attached shear layers in \fref{fig:f8}(a.1) and (a.2), respectively.
%Both elements present \oesim{-5}, comparable to the magnitudes seen in the instantaneous dissipation rate fields presented in \fref{fig:f5}(a).
Signatures of dissipation along the peripheries of inclined vortical structures are visible ($4 \le x/D < 8$) in both vertical and horizontal slices.
Non-trivial $\oek$ is also seen in the lee wave region above the seamount and the near wake.
When $\ui$ reaches zero at phase $\phi=3\pi/2$, \fref{fig:f8}(c.1) reveals clear signatures of the inclined layers corresponding to \oesim{-7} or greater.
Robust signatures of enhanced upstream dissipation are also seen in \fref{fig:f8}(c.2), owing to eddy-induced westward flow.

The $\oek$ field at $\phi = 2\pi$ in \fref{fig:f8}(d.2) is somewhat asymmetric, favoring larger values south of the topography, compared to the three other $\phi$.
This is likely related to the evolution of the vortex pairs that are mostly shed towards the south of the seamount, as the flow accelerates back up from $\ui=0$.
A well-formed near wake is absent during this interval and the effect of background rotation on the vortex pair generated at $\phi=3\pi/2$ is different depending on whether it is shed towards the north or south of the seamount.
It is possible a vortex pair is more likely to be shed south of the seamount due to the action of the Coriolis force, but this remains to be explored.
Nevertheless, the hydraulic jet and attached shear layers are seen again in \fref{fig:f8}(d), with \oesim{-7}, and signatures of nascent vertical overturnings are visible along inclined layers.

\begin{figure*}[!t]
    \centering
    \includegraphics[scale=0.714]{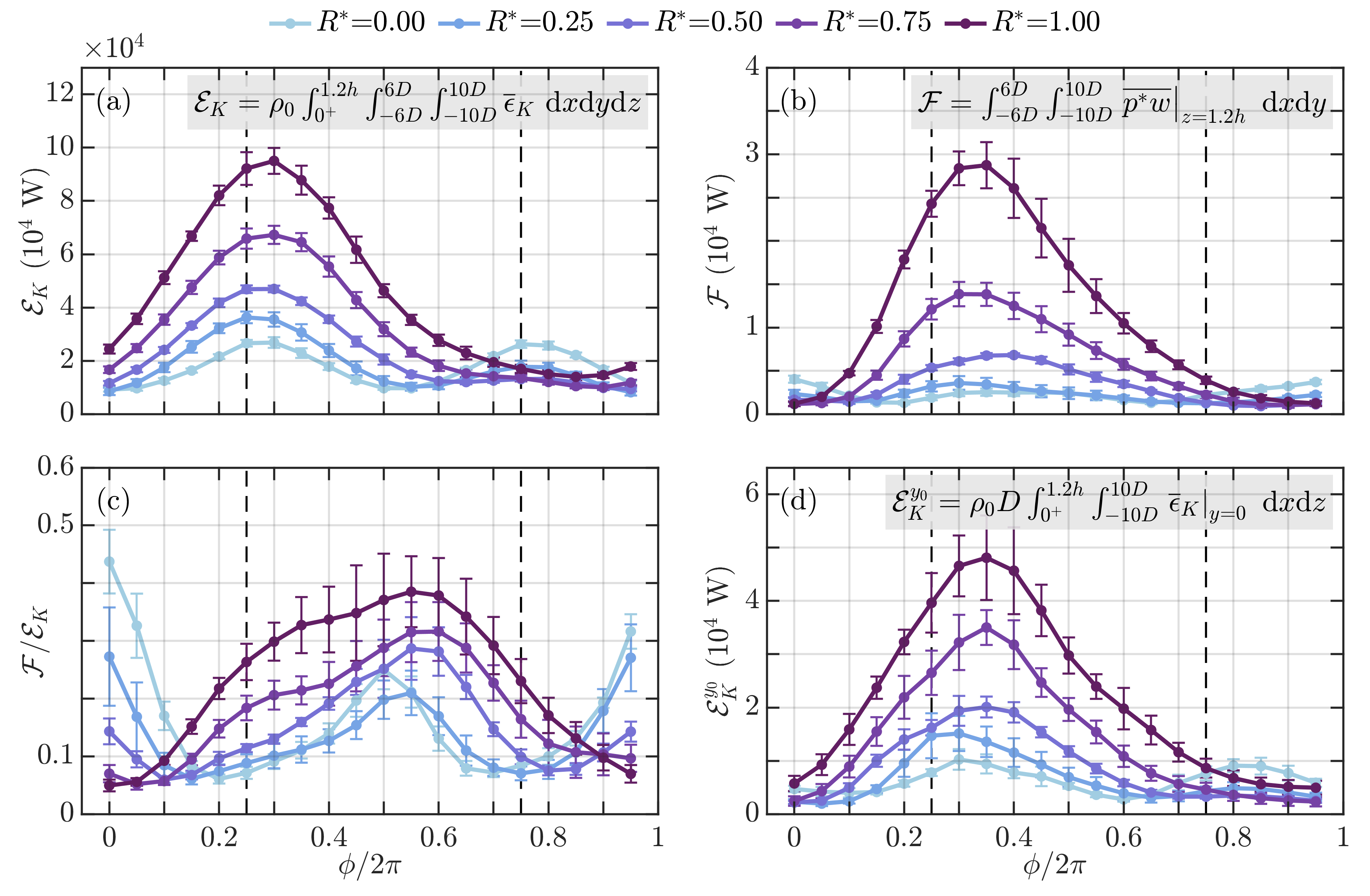}
    \caption{Top row shows the tidal phase dependence of: (a) the wake dissipation $\Ek$, and (b) the wave dissipation (i.e., the internal wave flux) $\F$.
    Bottom row: (c) $\F/\Ek$, the ratio of wave to wake dissipation, and (d) $\Ek^{y_0}$, an approximation of $\Ek$ obtained using an area integral over a vertical transect of measurements at $y/D=0$, as a function of $\phi$.
    The quantities in (a)--(d) are computed from phase-averaged fields, and the standard deviations across tidal cycles are represented using error bars at each $\phi$.
    The black dashed vertical lines correspond to $\phi=\pi/4$ and $3\pi/4$, the tidal phases when $\ui$ attains its maximum and minimum values.}
    \label{fig:f11}
\end{figure*}

To delineate the importance of the dissipation in the wake, we present the volume-integral of the kinetic energy dissipated in the wake of the seamount in \fref{fig:f11}.
A comparison is also made with the wave dissipation, i.e., kinetic energy transported away from the seamount by the internal wave flux, to potentially be dissipated elsewhere.
To understand how the wake dissipation from a bulk integral may be represented in field experiments, an area integral over a vertical transect as seen from a ship is also presented.

Figure~\ref{fig:f11}(a) presents the variation of the volume-integrated dissipation rate $\Ek$ (defined in the figure legend), as a function of $\phi$.
For $\rs=1$, $\Ek$ increases monotonically with $\phi$ during $0 \le \phi < \pi/2$, since newer and stronger dissipative structures are generated in the near wake as $\ui$ continues to accelerate.
$\Ek$ continues to increase as the $\ui$ starts to decelerate until $\Ek$ attains a peak shortly ($\approx37$ minutes for the $M_2$ tide in this problem) after $\phi=\pi/2$.
The offset is because of the persistence of strong dissipative features for a short while after their generation, while new dissipative features are generated.
For the remainder of the deceleration stage, from $\pi/2 < \phi \le \pi$, $\Ek$ decreases monotonically with $\phi$.
Similar trends are seen for all five $\rs$ for the first half of the tidal cycle, $0 \le \phi < \pi$.

As the tidal component starts flowing westward for $\phi>\pi$, $\Ek$ continues to decrease for $\rs=0.75$ and $1$, before increasing again towards the end of the cycle.
However, $\Ek$ for the three remaining $\rs=0$, $0.25$, and $0.5$, all of which correspond to $Ex^-(\rs)<0$ (the net current becomes westward), present a second peak in the latter half of the tidal cycle.
For all three cases, the second peak occurs near $\phi=3\pi/2$, at maximum westward current.
Interestingly, the $\Ek$ corresponding to the second maxima of $\rs=0$ and $0.25$, which also revealed strong peaks at $2f_t$ in the temporal spectral in \fref{fig:f4}(c) and (d), exceeds dissipation rates seen at the same phase for the other, larger values of $\rs$. The standard deviation in the volume integrals about $\Ek$ (plotted as error bars), which accounts for the variability across tidal cycles, is relatively small --- the maximum value is 6~Wm$^3$~kg$^{-1}$, for $(\rs,\phi)=(1,\pi/4)$, which is only 6.67\% of the corresponding value of $\Ek$.

%TC:ignore
%TC:endignore

Cycle-averaged $\Ek$, denoted as $\cav{\Ek}$, provides an overall space-time metric and is tabulated in the first row of Table~\ref{tab:diss}.
These values decrease monotonically as $\rs$ decreases from $1$ to $0$. The maximum current speed, $\um = \uc + \ut$, decreases systematically with decreasing $\rs$ as well, and this contributor to enhanced dissipation needs to be separated out for disentangling the role of the tide from the mean current. An appropriate normalization of cycle-averaged $\cav{\Ek}$ is the inviscid scale ($U^3/L$) of dissipation rate multiplied by a scale for the wake volume ($hD^2$) and the reference density ($\rho_0$). Such normalization using $U=\um$ and $L=D$ leads to the fourth row of Table~\ref{tab:diss}. The non-dimensionalized $\cav{\Ek}$ {\em increases} as the tide becomes more dominant, which is the opposite of the trend in dimensional $\Ek$. \textit{The increase is three times for the pure tide with respect to  $\rs=1$}. It is a remarkable consequence of the complex eddy-eddy and eddy-wake-topography interactions (described in previous sections) that come into play when the current reverses direction, a facet of tidally dominated flows at 3D topography.

%TC:ignore
\begin{figure*}[!t]
    \centering
    \includegraphics[scale=0.714]{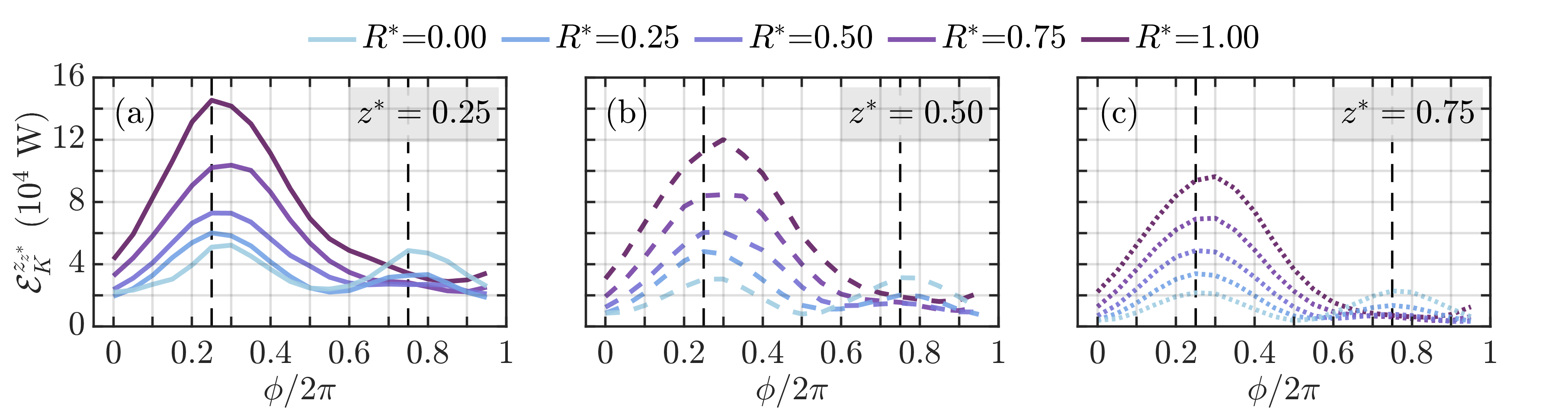}
    \caption{Phase dependence of $\Ek^{z_{\zs}}$, obtained using a horizontal-area integral, at three different heights $\zs=z/h$: (a)~$0.25$, (b)~$0.50$, and (c)~$0.75$.}
    \label{fig:f11_1}
\end{figure*}
%TC:endignore

%TC:ignore
\begin{table*}[!t]
\centering
\begin{NiceTabular}{|W{c}{3.2cm} W{c}{1.6cm}||W{c}{1.4cm}|W{c}{1.4cm}|W{c}{1.4cm}|W{c}{1.4cm}|W{c}{1.4cm}|}%[hvlines]
    \CodeBefore
    % \cellcolor[HTML]{DCDCDC}{1-,-1,-2}
    % \cellcolor[RGB]{220,220,220}{1-,-1,-2}
    % \cellcolor[RGB]{240,240,240}{3-7,4-7}
    % Removed shading according to JPO guidelines; added double lines instead
    \Body
    \Hline
    \Block{1-2}{\diagbox{~\hspace{0.75em}Quantities\vspace{0.3em}}{\vspace{0.5em}$\rs$\hspace{1em}~}}& & 1.00 & 0.75 & 0.50 & 0.25 & 0.00 \\
    \hline\hline
    $\cav{\Ek}$&(kW) & 45.24 & 32.00 & 22.83 & 18.50 & 17.46 \\
    \hline
    $\cav{\F}$&(kW) & 11.54 & 5.89 & 3.26 & 2.15 & 2.38 \\
    \hline
    $\cav{\F}\big/\cav{\Ek}$&(\%) & 25.50 & 18.42 & 14.30 & 11.64 & 13.61 \\
    \hline
    {$\cav{\Ek}\big/ \rho_0\um^3hD$} & -- & 0.075 & 0.080 & 0.090 & 0.126 & 0.232 \\
    \hline
 %  {$\cav{\Ek}\big/ \rho_0\um^2NhD^2$} & -- & 0.0068 & 0.0063 & 0.0061 & 0.0071 & 0.0105 \\
  %  \hline
    $\cav{\Ep}\big/\big(\cav{\Ep}+\cav{\Ek}\big)$& -- & 0.154 & 0.160 & 0.163 & 0.165 & 0.181 \\
    \hline
    $\cav{\Ep}\big/\cav{\Ek}$& -- & 0.182 & 0.191 & 0.194 & 0.197 & 0.221 \\
    \Hline
\end{NiceTabular}
\vspace{0.5em}
\caption{Tidal-cycle averages of various quantities, defined as $\cav{\R{\xi}}=\int_{0}^{2\pi}\R{\xi}~\mathrm{d}\phi\big/2\pi$, for the five values of $\rs$ considered.}
\label{tab:diss}
\end{table*}
%TC:endignore

The internal wave flux $\F$ for $\rs=0.75$ and $1$ in \fref{fig:f11}(b) presents a peak shortly ($\approx75$ minutes for an $M_2$ tide) after $\phi=\pi/2$ due to the persistence of the large amplitude lee waves generated when $\ui=\uc+\ut$.
As was seen for $\Ek$, the $\rs\le0.5$ cases with $Ex^-(\rs)<0$ present more than one local maxima in $\F$.
From Table~\ref{tab:diss}, the average of $\F$ over the tidal cycle also decreases from $\rs=1$ to $0.25$, but increases slightly when $\rs$ becomes 0.
%potentially due to the action of internal tides.
As a result, the ratio of cycle-averaged $\F$ and $\Ek$ also decreases from 25.5\% at $\rs=1$ to 11.64\% at $\rs=0.25$, before increasing to 13.61\% at $\rs=0$ (Table~\ref{tab:diss}).
This is somewhat evident from \fref{fig:f11}(c), wherein the $\F/\Ek$ values for $\rs=0$ are close to or greater than those for $\rs=0.25$.
Additionally, $\F/\Ek<1$ throughout the tidal cycle ($0 \le \phi < 2\pi$), for all five values of $\rs$.

Finally, to emulate observational measurements from a boat during field experiments, \fref{fig:f11}(d) presents $\Ek^{y_0}$, obtained from dissipation rate measurements on a vertical transect passing through ${y/D=0}$. The integral over the transect area is multiplied by $D$ (assumed lateral scale for the wake) to enable direct comparison of $\Ek^{y_0}$ and $\Ek$.
The $\Ek^{y_0}$ curves are qualitatively similar to the $\Ek$ curves in \fref{fig:f11}(a), but with the peaks offset further to the right of $\phi=\pi/2$.
This is because the vertical transect fails to sample the strong dissipation rates associated with the attached lateral shear layers, which
are strongest when the current $\ui$ peaks.
Additionally, the magnitude of $\Ek^{y_0}$ is weaker than $\Ek$ at all tidal phases, for all values of $\rs$, and on average, the ratio ${\Ek/\Ek^{y_0}\approx3}$.
The error bars are also smaller, but the maximum value of 0.016~Wm$^2$~kg$^{-1}$ is 18\% of the corresponding $\Ek^{y_0}$ value.

%TC:ignore
\begin{figure*}[!t]
    \centering
    \includegraphics[scale=0.714]{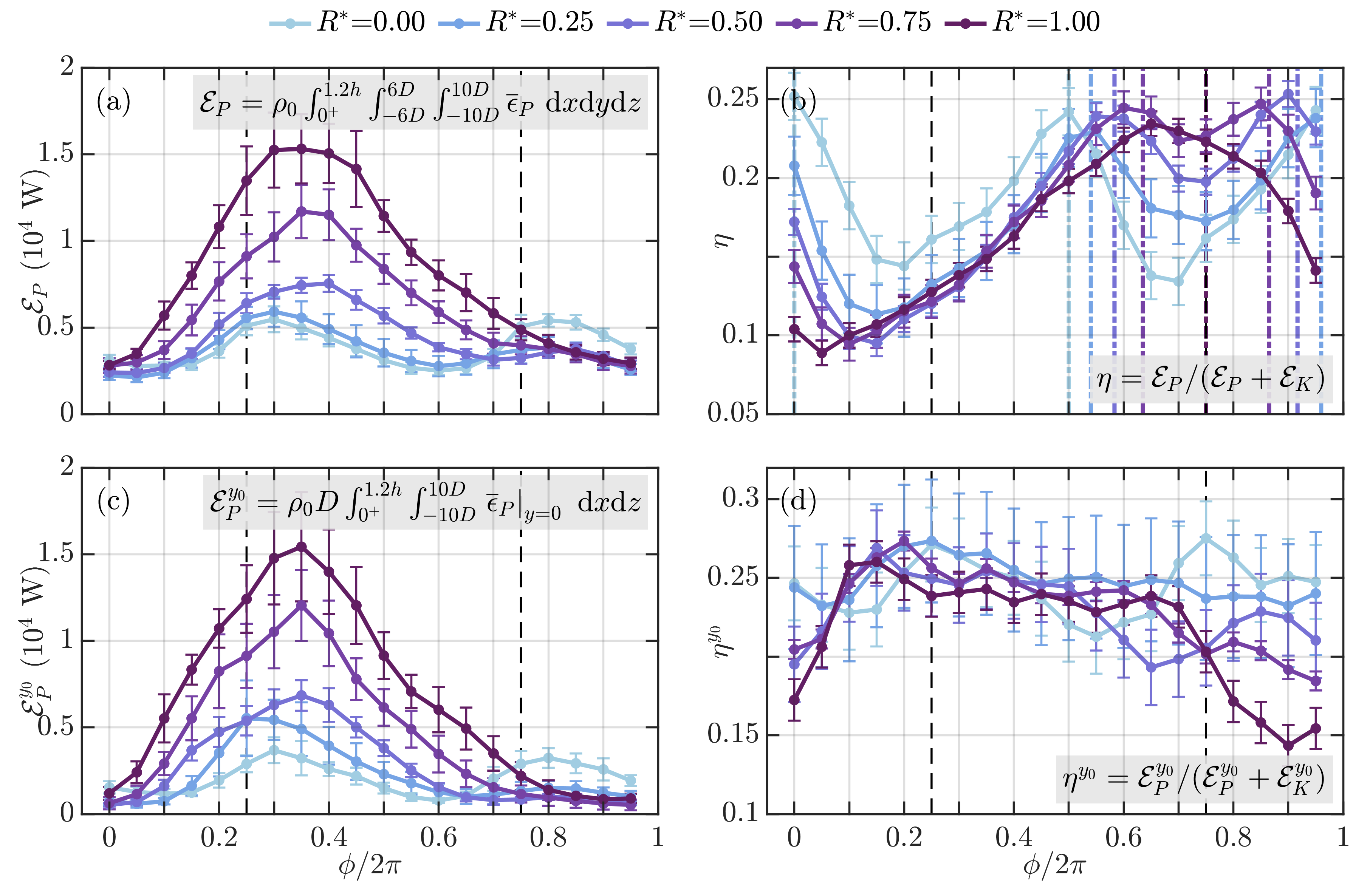}
    \caption{Top row shows phase dependence of: (a) the wake dissipation of available potential energy per unit mass $\Ep$, and (b) the bulk mixing efficiency $\n$.
    Bottom row: phase dependence of the approximations based on a streamwise vertical transect (c) $\Ep^{y_0}$, and (d) $\n^{y_0}$, of the quantities on the top.
    Standard deviations across tidal cycles are represented using error bars at each tidal phase.
    $\phi=\pi/4$ (maximum $\ui$) and $3\pi/4$ (minimum $\ui$) are shown using black dashed vertical lines and the zero-$\ui$ phase by lines in color (depends on $\rs$).}
    \label{fig:f12}
\end{figure*}
%TC:endignore

In the ocean, horizontal transects can also be sampled using autonomous gliders.
The vertical variation of dissipation rates in the wake is nevertheless of interest, given the complex vortex dynamics observed in the wake (see \fref{fig:f2} and \fref{fig:f3}, for example).
The quantity $\Ek^{z_\zs}$, which is the horizontal-area integral of $\oek$ multiplied by $h$ and $\rho_0$, computed at three different $\zs=z/h=0.25$, $0.5$ and $0.75$ are presented in \fref{fig:f11_1}(a), (b) and (c), respectively.
The $\Ek^{z_\zs}$ curves for all five $\rs$ are qualitatively similar to each other (and the bulk $\Ek$ in \fref{fig:f11}a), despite the vertical variability in wake-vortex dynamics.
The magnitudes of $\Ek^{z_\zs}$ decrease monotonically with increasing $\zs$, likely linked to the decrease in the horizontal size of the vortical structures with $\zs$, but a clear scaling with the local diameter $d_\zs=(1-\zs)D$ is not found.

When investigating ocean mixing, the quantity of direct interest is the dissipation rate of the available potential energy per unit mass $\ep$.
The spatial distribution of $\ep$ is related to the distribution of $\ek$ (c.f. Supplemental Fig.~S3 and \fref{fig:f5}).
The $\ep$ is maximized in regions where nearby fluid is `entrained' and mixed, which usually lies close to the regions of strong shear, which in turn are regions of enhanced $\ek$.
The volume integral of $\ep$ in the wake region, $\Ep$, is presented as a function $\phi$ in \fref{fig:f12}(a).
The $\Ep$ curves are qualitatively similar to the $\Ek$ in \fref{fig:f11}(a), albeit with lower magnitudes and flatter peaks offset to the right of $\phi=\pi/4$.
The error bars in $\Ep$, as a fraction of the corresponding $\Ep$, have larger magnitudes in \fref{fig:f12}(a) compared to \fref{fig:f11}(a).

The variation of the bulk mixing efficiency $\n$ for the wake region, as a function of $\phi$, is presented in \fref{fig:f12}(b).
As the tidal component accelerates from zero, $\n$ first decreases to its minimum, before increasing with $\phi$.
This behavior is likely because $\Ek$ initially increases faster than $\Ep$ as the shear increases, before the shear instabilities grow to finite amplitude and start mixing the fluid, when $\Ep$ takes over.
The minimum in $\n$ occurs earlier and becomes progressively smaller as $\rs$ is increased --- the greater the $\uc=\ut\rs$, the stronger the shear, and earlier the shear instabilities grow to finite amplitude.
As the flow evolves and vortical structures become better developed, the rate of entrainment and mixing ($\Ep$) increases faster than $\Ek$ for all $\rs$ until $\n$ attains a local maximum.
Interestingly, the maximum $\n$ is attained near $\ui=0$ for all $\rs$ --- this is because the flow features that result in the strongest $\ek$, i.e., the hydraulic jet and attached shear layers, are absent when $\ui=0$, while the existing turbulence continues to mix the buoyancy gradients created during the preceding energetic phase of the current.
The $\n$ curve decreases with $\phi$ after the local maximum, before increasing again to a second maximum for $\rs<1$ at a phase $\phi>3\pi/2$, when $\ui$ again becomes zero as the current reverts to an eastward flow.
Interestingly, a larger $\rs$ corresponds to a shorter time interval between the zero-crossings of $\ui$, a greater second minimum of mixing efficiency $\n$, and a lesser first minimum of $\n$.

Within a tidal cycle, the value of $\n$ varies between 0.1 and a maximum of approximately $0.25$ for all five $\rs$, which corresponds to standard deviations of 20--30\% about the mean value $\cav{\eta}$ for different $\rs$.
An average bulk mixing efficiency for the tidal cycle, defined as $\cav{\Ep}\big/\big(\cav{\Ep}+\cav{\Ek}\big)$, is presented in the fifth row of Table~\ref{tab:diss}.
This average $\n$ follows the trend of the non-dimensionalized $\cav{\Ek}$ --- it {\em increases} as the tide becomes more dominant, with a minimum of 0.154 for $\rs=1$, and a maximum of 0.181 for $\rs=0$.
The corresponding values of an average mixing coefficient is presented in the last row, and the values all lie close to the widely used value of $0.2$.

If only a vertical transect of measurements is available, the approximation $\Ep^{y_0}$ in \fref{fig:f12}(c) is qualitatively similar to $\Ep$.
For the available potential energy, the ratio $\Ep/\Ep^{y_0}\approx 2$ (averaged across $\rs$ and all $\phi$) is smaller than the ratio for the kinetic energy.
The mixing efficiency calculated from the vertical transect $\n^{y_0}$ in \fref{fig:f12}(d), however, is markedly different from the bulk mixing efficiency $\n$.
The $\n^{y_0}$ magnitudes for all $\rs$ remain close to 0.26 for the majority of the tidal cycle, and the standard deviations for the five $\rs$ are only 5--17\% about $\cav{\n^{y_0}}$.
The low values of $\n$ seen during the eastward (same direction as mean) tide, especially as the flow accelerates, are not seen in the centerline-transect value $\n^{y_0}$.
%and do not reflect the peaks near zero-crossings of $\ui$ seen for $\n$ in \fref{fig:f12}(b).
The error bars for $\n^{y_0}$ at each tidal phase in \fref{fig:f12}(d), however, are larger, indicating its values likely fall in the range $0.2\le\n^{y_0}\le0.3$ for the majority of the tidal cycle, $\pi/10\le\phi\le3\pi/2$.

\section{Discussion and Conclusions}
\label{sec:disc}
Large eddy simulations (LESs) were performed to investigate the vortex dynamics, turbulence, dissipation, and mixing in the wake of {\em tidally dominated} currents past three-dimensional (3D) topography in a linearly stratified environment with weak background rotation.
The barotropic flow is modeled as $\ui(t)=\uc+\ut\sin(2\pi f_t t)$, where $\uc$ is the mean current and $\ut$ is the amplitude of the tidal oscillation, and the relative strength of the tidal flow is quantified using the ratio $\rs=\uc/\ut$.
To investigate tidally dominated flows, $\rs$ was varied from $0$ to $1$ by varying $\uc$ from $0$ to $\ut$.
The idealized 3D topography, a conical seamount with height $h$ and base diameter $D$, is supercritical, and a hydraulic response was observed near the apex.
The seamount is also dynamically tall, with the topographic Froude number $Fr_h\le0.3$ across $\rs$, and lateral flow separation resulted in the generation of wake eddies.

Investigation of the flow evolution within a tidal cycle revealed lee waves emanating from the top of the seamount, a wake region below, and a thin hydraulic jet near the seamount at the interface between the two.
The wake region consists of a turbulent recirculation region close to the seamount (`near wake') enclosed laterally within attached shear layers, and \R{vertically coupled} coherent vortices shed and advected away from the near wake (`wake eddies/vortices').
In the $\rs=1$ simulation, a Strouhal pair of coherent vortices was shed during the first half of the tidal cycle, and a vortex pair was shed during the second.
For a weaker mean current ($\rs=0.5$), only one coherent vortex was shed in the first half of the cycle before a vortex pair was generated during the latter half.
The wake vortex dynamics in the absence of a mean current ($\rs=0$) was significantly different due to the purely oscillatory $\ui$.
In the LES, a coherent vortex is shed to the north of the topography during the first half of the tidal cycle.
During reverse flow in the next half, the vortex is advected back towards the seamount, and its interaction with the recirculation region results in a vortex of the opposite polarity shed to the north on the other side of the seamount.
Since the simulations are performed on an $f$-plane, the consistent northward shedding is likely one state of a bistable system.

\R{For reference, simulations performed for $\rs=0$ with $\ui(t)=\ut\sin(2\pi f_t t + \pi)$, i.e., a phase difference of $\pi$ for the tidal oscillation, revealed consistent southward shedding (not shown).
Systems are often bistable on timescales that are orders of magnitude longer than any convective timescale~\citep{williams2025asymmetries}.
Hence, the system presented here could potentially switch to southward shedding after a sufficiently long time interval, or due to the action of a sufficiently large-amplitude perturbation.}

Spectral analysis of the streamwise velocity measured at a point within the near wake revealed a dominant peak at the tidal frequency $f_t$ for all five $\rs$.
The dominant frequencies in the spanwise velocity spectra, however, were not all at $f_t$, suggesting tidal synchronization of the near-wake dynamics to frequencies other than $f_t$, as observed for the tidally modulated cases ($\rs > 1$) in~\citet{puthan2022wake}.
The horizontal-area-integrated vertical vorticity spectra reflected the tidal synchronization of wake vortices: $\rs=1$, $0.75$, and $0.5$ had dominant peaks at $f_t$, whereas the two tidally most dominant cases, $\rs=0$ and $0.25$, had them at $2f_t$.
The spectra of horizontal-area-integrated dissipation rates of kinetic energy per unit mass $\ek$ were qualitatively similar to the vorticity spectra.

The wake dissipation exhibited large values (\esim{-6.5}) at the coherent vortex peripheries across $\rs$. The dissipation rate field is spread out and patches with \esim{-6.5} are encountered laterally up to $2.5D$ from the topography owing to vortex-vortex interactions. The lateral spread is larger for a purely tidal flow.  Owing to the mean advection, patches with  \esim{-6.5} are found far downstream, e.g., $6D$ downstream of the seamount.
Stronger dissipation  (approximately an order of magnitude greater) was observed in the near wake with $x/D < 3$ in all cases, and the largest values  (\esim{-5}) were consistently observed in the hydraulic jet and attached shear layers. The largest $\ek$ regions, being highly localized,  are difficult to sample in observational transects and present a challenge when obtaining field estimates of wake dissipation.

The volume-integrated wake dissipation $\Ek$ peaked shortly after maximum $\ui$ (tidal phase $\pi/2$) for all five $\rs$, and was consistently greater than the internal wave flux --- by a factor of about 4 at $\rs = 1$ to about 7 at $\rs = 0$.
Thus, the local dissipation in this 3D topographic wake was substantially greater than the energy that is transported away to potentially be dissipated elsewhere.
The magnitudes of the horizontal-area integrals of dissipation rates within the wake decreased with increasing height above the bottom, but their tidal phasing remained qualitatively similar throughout the tidal cycle.
The tidal-cycle-averaged dissipation rate magnitudes decreased with $\rs$ in the simulations, but their normalized values were the greatest for tidally the most dominant cases.
This is an important overall consequence of the complex eddy-eddy and eddy-wake-topography interactions that occur as the flow reverses direction.

The tidal phasing of the dissipation rate of available potential energy in the wake $\Ep$ shared some similarities with $\Ek$.
The magnitudes of $\Ep$ attained a maximum after peak $\ui$, but the peaks were offset further from phase $\pi/2$ compared to $\Ek$.
The bulk mixing efficiency $\n$ varied between 0.1 and 0.25 within a tidal cycle.
Interestingly, $\n$ was maximized near $\ui=0$, for all $\rs$.
This is because the hydraulic jet and attached shear layers, which contribute significantly to $\ek$, are absent when $\ui=0$, while the existing turbulence continues to mix the buoyancy gradients created during the preceding energetic phase of the current.

Observationally, it is often easier to obtain vertical transects of measurements from a boat, compared to full three-dimensional fields.
Analysis of the simulation data revealed that bulk mixing efficiencies obtained using measurements from a vertical transect alone differ from the bulk values.
However, the approximations of volume-integrated dissipation rates of kinetic energy and available potential energy, obtained from a vertical transect passing through ${y/D=0}$, were qualitatively similar to the bulk values.
Quantitatively, these approximations differed by a factor of 2--3 on average throughout the tidal cycle, for all $\rs$.
A closer investigation revealed that the vertical shear components $S_{13}$ and $S_{23}$ are the major contributors to $\ek$.
In vertical transects, $S_{13}$ and $S_{23}$ resulted in 89\% or more of the phase-averaged dissipation rates of kinetic energy at all tidal phases, across the five $\rs$ considered (not shown).

When drawing parallels between observations and the results presented here, it is important to remember that the non-dimensional groups $Re$, $Ro$, and $Fr_h$ presented are all defined using the velocity at peak tidal flow, $\um$.
These metrics should be associated with the bulk dynamics over an entire tidal cycle, and {\em not} the local dynamics at an instant.
The latter depends on the instantaneous $\ui$, which can lie in the range $0\le|\ui|\le\um$ for the tidally dominated cases investigated here.
Other non-dimensional groups like the gradient Richardson number, vorticity Rossby number, and the angle $\phi_{Ri_B}$ related to the potential vorticity introduced by \citet{thomas2013symmetric}, for example, are pertinent to local investigations of fluid instabilities.
\R{A closer investigation of the instabilities that lead to turbulence in the near wake, coherent eddies, hydraulic jet, and attached shear layers is of future interest.}

Additionally, drawing inferences about the bulk, based on the instantaneous values of non-dimensional parameters alone, is inappropriate.
For example, $(\rs,\phi)=(0,\pi)$ and $(\rs,\phi)=(1,3\pi/2)$ have the same values for $Re$, $Ro$, and $Fr_h$ defined using instantaneous $\ui$, but correspond to very different wakes, and hence cannot be compared.
{ The wakes are qualitatively different for the same values of instantaneous non-dimensional groups for the same $\rs$ --- $(\rs,\phi)=(1,\pi)$ and $(\rs,\phi)=(1,2\pi)$, for example --- and hence, one needs to account for the tidal phasing of the flow.}

Finally, the idealized large eddy simulations presented here considered a barotropic inflow in a linearly stratified environment.
The barotropic tide is modeled as a sinusoidal wave oscillating at a characteristic tidal frequency $f_t$, with the ratio $\rs$ determining whether the flow changes direction within a tidal cycle.
While this is a reasonable approximation for diurnal and semi-diurnal tides, it is not suitable for mixed tides.
The spring-neap modulation of tidal flows can also have significant effects on the observed wake dynamics and mixing.
A more realistic baroclinic current in a non-linear stratification can result in modified wake dynamics, depending on the pycnocline thickness and its relative position vertically with respect to the topography, for example, and requires further investigation.

% \section*{Bookmark}

% \clearpage
%%%%%%%%%%%%%%%%%%%%%%%%%%%%%%%%%%%%%%%%%%%%%%%%%%%%%%%%%%%%%%%%%%%%%
% ACKNOWLEDGMENTS
%%%%%%%%%%%%%%%%%%%%%%%%%%%%%%%%%%%%%%%%%%%%%%%%%%%%%%%%%%%%%%%%%%%%%
\acknowledgments
This work was supported by the ONR grant N00014-23-1-2170, and high-performance computer time and resources from the DoD High Performance Computing Modernization Program.
The authors declare that they have no known competing financial interests or personal relationships that could have appeared to influence the work reported in this paper.

%TC:ignore
%%%%%%%%%%%%%%%%%%%%%%%%%%%%%%%%%%%%%%%%%%%%%%%%%%%%%%%%%%%%%%%%%%%%%
% DATA AVAILABILITY STATEMENT
%%%%%%%%%%%%%%%%%%%%%%%%%%%%%%%%%%%%%%%%%%%%%%%%%%%%%%%%%%%%%%%%%%%%%
\datastatement
The datasets generated and/or presented can be accessed at the following DOI: \href{https://zenodo.org/records/15391881}{10.5281/zenodo.15391881}.

%%%%%%%%%%%%%%%%%%%%%%%%%%%%%%%%%%%%%%%%%%%%%%%%%%%%%%%%%%%%%%%%%%%%%
% REFERENCES
%%%%%%%%%%%%%%%%%%%%%%%%%%%%%%%%%%%%%%%%%%%%%%%%%%%%%%%%%%%%%%%%%%%%%
\bibliographystyle{ametsocV6}
\bibliography{references}

%TC:endignore
\end{document}